\documentclass[final,authoryear,12pt,5p,times]{elsarticle}

\usepackage{bm}

\newcommand{\bmath}[1]{\bm{#1}}

\newcommand{\mathbmss}[1]{\bm{\mathsf{#1}}}
\newcommand{\var}{\mathop{\text{Var}}\nolimits}
\newcommand{\cov}{\mathop{\text{Cov}}\nolimits}

\newcommand{\expect}{\mathop{\mathbb{E}}}
\newcommand{\disp}{\mathop{\mathbb{D}}}
\newcommand{\As}{\mathop{\mathbb{A}\text{s}}}

\newcommand{\diag}{\mathop{\text{diag}}\nolimits}
\newcommand{\const}{\mathop{\text{const}}\nolimits}
\newcommand{\FAP}{{\text{FAP}}}
\newcommand{\Vol}{{\text{Vol}}}

 \usepackage{graphicx}

\usepackage{amssymb}
\usepackage{amsmath}

\journal{Astronomy and Computing}

\begin{document}
\sloppy

\begin{frontmatter}



\title{Statistical detection of clumps and gaps in a random sample by continuous wavelet
transforms: the 2D case}


\author[GAO,SPbSU]{Roman V. Baluev}
 \ead{r.baluev@spbu.ru}

\author[SPbSU]{Evgeniy I. Rodionov}
\author[SPbSU]{Vakhit Sh. Shaidulin}

\address[GAO]{Central Astronomical Observatory at Pulkovo of Russian Academy of Sciences,
Pulkovskoje sh. 65/1, St Petersburg 196140, Russia}
\address[SPbSU]{Saint Petersburg State University, Faculty of Mathematics and Mechanics, Universitetskij
pr. 28, Petrodvorets, St Petersburg 198504, Russia}

\begin{abstract}
We present a self-consistent framework to perform the wavelet analysis of two-dimensional
statistical distributions. The analysis targets the 2D probability density function
(p.d.f.) of an input sample, in which each object is characterized by two peer parameters.
The method performs a probabilistic detection of various `patterns', or `structures'
related to the behaviour of the p.d.f. Laplacian. These patterns may include regions of
local convexity or local concavity of the p.d.f., in particular peaks (groups of objects)
or gaps. In the end, the p.d.f. itself is reconstructed based on the least noisy (most
economic) superposition of such patterns. Among other things, our method also involves
optimal minimum-noise wavelets and minimum-noise reconstruction of the distribution density
function.

The new 2D algorithm is now implemented and released along with an improved and optimized
1D version. The code relies on the C++11 language standard, and is fully parallelized.

The algorithm has a rich range of applications in astronomy: Milky Way stellar population
analysis, investigations of the exoplanets diversity, Solar System minor bodies statistics,
etc.
\end{abstract}


\end{frontmatter}



\section{Introduction}
The wavelet analysis, ascending in its early form to \citet{GrMorlet84}, is a mathematical
tool that allows to decompose a function into multiple levels of different resolutions.

Nowdays, the wavelet analysis technique is frequently used in various fields of astronomy.
The single-dimensional version is useful in the time series analysis, because it allows to
track the time evolution of quasi-periodic variations \citep{Foster96c}, while
two-dimensional wavelets are often used in the image analysis.

However, most of these classic applications of wavelets target functions of some
deterministic argument $x$ (where $x$ can be from $\mathbb R^n$ in general). Although the
function itself, $f(x)$, may contain random noise, this function is assumed to be defined
on some continuous domain or on a fine enough grid to approximate this continuous domain.
In this case we just measure the values $f_i=f(x_i)$ and apply a discretized version of the
wavelet transform to these values. But in this work we consider the application of the
wavelet technique to a not so popular but still practically important task: analyse a
probability density function $f(x)$ of some random variable, or a pair of (possibly
correlated) random variables. In this case we do not measure any $f_i$ at all. Instead, we
should estimate the wavelet transform based on some finite-size statistical sample
$\{x_i\}_{i=1}^N$. Here the function $f(x)$ is determined implicitly, as a local density of
these $x_i$ in the vicinity of the specified point $x$. As we can see, the mathematical
formulation of this task is qualitatively different from what is used in the classic
wavelet analysis.

The single-dimensional case of this task was already considered in \citep{Baluev18a}, with
an application to exoplanetary distributions presented in \citep{Baluev18b}. In this work,
we present an extension of the 1D theory to the case of two dimensions. The 2D case is even
more practical perhaps, since it considerably more informative than the 1D one: various
statistical families can often be resolved from each other in 2D, even if they overlap in
the both 1D distributions. Therefore, 2D analysis of statistical samples can be useful in
multiple fields: Milky Way population studies, especially in view of the emerging GAIA data
\citep{Gaia18}, statistical and dynamical studies of asteroids in Solar System, studies of
exoplanetary statistics, which now became quite reach thanks to the ground-based as well as
space projects like Kepler and the ongoing TESS.

Unfortunately, there is only a relatively meagre amount of literature available with
respect to the analysis of statistical distrubutions by wavelets. First attempts to apply
the wavelet transform technique to reveal clumps in stellar distributions date back to
1990s \citep{Chereul98,Skuljan99}, and \citet{Romeo03,Romeo04} suggested the use of
wavelets for denoising results of $N$-body simulations. Nowdays, wavelet transforms are
quite routinely used to analyse CMB data from WMAP \citep{McEwen04,McEwen17}. The very idea
of `wavelets for statistics' from the mathematical point of view was also considered by
\citet{Fadda98,ABS00}.

But after a closer look, many of these wavelet analysis methods appeared deficient in some
aspects. This discussion is given in details in \citep{Baluev18a}, and here we highlight
only two main issues. The first one is that the wavelet transform is often applied to a
binned sample. This allows to use merely the classic wavelet analysis technique with only
minor modifications concerning the noise treatment (the noise in each bin or 2D box appears
Poisson rather than Gaussian). However such an approach implies obvious caveats: the
binning always leads to (i) loosing the small-scale resolution and (ii) additional
interpolation-like errors. Yet another issue appears when thresholding the noise in the
wavelet transform. To perform such a thresholding, we obviously must determine some
significance level for wavelet coefficients. However, since in practice we always consider
a large domain in the shift-scale space rather than just a single point, it is necessary to
consider the significance of multiple coefficients at once, rather than each individually.
This means that the significance levels must be determined based on the corresponding
extreme value distribution (EVD), rather than single value distribution (SVD). This issue
has often been ignored in the literature, e.g. by \citet{Skuljan99}, but then our analysis
result may occur to depend on a single outlier wavelet coefficient, which become
increasingly more frequent if we expand the analysis domain. Another issue with the
significance levels appear because they are usually computed from Monte Carlo simulations,
which is a very slow and heavy computation. The need of analytic significance formulae
therefore becomes obvious.

In this paper we mainly present the mathematical theory from \citep{Baluev18a} extended to
the 2D case, and describe our new optimized computating code (both for the 1D and 2D
cases). Below we also provide a quick 2D test case to demonstrate the practical use of our
algorithm and its possible benefits. However, in this paper we did not intend to do a
comprehensive `field-testing' of the code or to apply it to a particular sample of objects.
The latter task is left for future work.

The code is available for download at
\texttt{https://sourceforge.net/projects/waveletstat/}. It is open-source and licensed
under the MIT license. The download archive also contains an 1D sample test case
application (sample data, code invokation script, and {\sc GNUPLOT} script to render the
results in graphical form)

\section{Overview of the paper}
\label{sec_overview}
In this paper we undertook an extensive work following generally the same sequence as in
the 1D case \citep{Baluev18a}.
\begin{enumerate}
\item We start from presenting some basic mathematical definitions, general discussion and
justification of the method (Sect.~\ref{sec_basic}).
\item Further on, we present a self-consistent statistical theory, including the
sample-based estimation of the Continuous Wavelet Transform (or CWT), construction of the
statistical tests and analytic calibration of the CWT noise levels and its normality
(Sect.~\ref{sec_detect}).
\item Then we find optimal wavelets that provide minimum noise, simultaneously constraining
the CWT noise at a high order of normality, as well as an optimal minimum-noise
reconstruction kernel for the inverse CWT (Sect.~\ref{sec_opt}).
\item We discuss an iterative matching pursuit like algorithm that builds a p.d.f. model
based on CWT noise thresholding (Sect.~\ref{sec_recon}).
\item In Sect.~\ref{sec_code}, we discuss our updated 1D and new 2D code that efficiently
implement all this theory.
\item In Sect.~\ref{sec_tests} we present some numeric tests to verify the statistical
validity of our 2D algorithm, and also a demonstrative example exposing an analysing of a
2D distribution of the Main belt asteroids.
\end{enumerate}

\section{Wavelet transforms in two or more dimensions: basic mathematics}
\label{sec_basic}
\subsection{Definitions of the CWT in multiple dimensions}
The first definition of the CWT uses the shift (or location) $n$-dimensional vector $\bmath
b$ and the $n\times n$ scale matrix $\mathbmss A$. We will call it the $Y$-shape:
\begin{equation}
Y(\mathbmss A,\bmath b) = \int f(\bmath x) \psi\left(\mathbmss A^{-1} (\bmath x-\bmath b)\right) d\bmath x.
\label{ycwt}
\end{equation}

Alternatively, the so-called $\Upsilon$-shape might be convenient sometimes:
\begin{equation}
\Upsilon(\mathbmss K,\bmath c) = \int f(\bmath x) \psi\left(\mathbmss K \bmath x + \bmath c\right) d\bmath x.
\label{ucwt}
\end{equation}
It involves the phase-like vector argument $\bmath c$ and the wavenumber-like matrix
argument $\mathbmss K$.

These $Y$-shape and $\Upsilon$-shape formulae define exactly the same quantity,
parametrized by different systems of independent variables. These variables are mutually
bound as:
\begin{equation}
\mathbmss A = \mathbmss K^{-1}, \quad \bmath c = - \mathbmss K \bmath b.
\end{equation}
This assumes that $\mathbmss K$ is always non-degenerate, so that $\mathbmss A$ exists. All
the quantities, vectors andd matrices above are assumed real.

We can alternatively rewrite the $Y$-shape as
\begin{equation}
Y(\mathbmss A, \bmath b) = |\det \mathbmss A| \int f(\bmath b + \mathbmss A \bmath t) \psi(\bmath t) d\bmath t.
\label{ycwt2}
\end{equation}
If $\psi$ is well-localized, which is true for what we call `wavelet', then such $\psi$
basically cuts in $f(x)$ a domain centered at $\bmath b$ and sized and shaped according to
the scale parameters from the matrix $\mathbmss A$.

This very general definition of the CWT implies $n+n^2=n(n+1)$ degrees of freedom. Already
in the 2D case, $n=2$, this results in $6$ degrees of freedom, which is probably too large
to be practical.

Now let us assume that our wavelet is radially symmetric:
\begin{equation}
\psi(\bmath t) = \psi(t).
\end{equation}

Then let us write
\begin{equation}
\bmath t^2 = (\bmath x-\bmath b)^{\rm T} (\mathbmss K^{\rm T} \mathbmss K) (\bmath x-\bmath b).
\end{equation}
As we can see, in this case the CWT depends on the real symmetric positive-definite matrix
$\mathbmss K^{\rm T} \mathbmss K$, implying the reduction of total number of degrees of
freedom from $n(n+1)$ to $n+n(n+1)/2 = n(n+3)/2$.

Since $\mathbmss K^{\rm T} \mathbmss K$ is real symmetric positive-definite matrix, let us
diagonalize it:
\begin{equation}
\mathbmss K^{\rm T} \mathbmss K = \mathbmss R^{\rm T} \mathbmss D \mathbmss R, \quad \mathbmss D = \diag(\kappa_i^2), \, \quad \kappa_i = 1/a_i.
\end{equation}
Here, $\mathbmss R$ is an orthogonal matrix. Such an $\mathbmss R$ may describe, in
general, any isometry transform in $\mathbb R^n$ that includes e.g. rotations, reflections,
and permutations. However, reflections and permutations do not provide any useful new
information, so we restrict $\mathbmss R$ to be a rotation matrix.

Then, without loss of generality, we may set
\begin{equation}
\mathbmss K = \mathbmss D^{1/2} \mathbmss R, \quad \mathbmss A = \mathbmss R^{\rm T} \mathbmss D^{-1/2}, \quad \bmath c = -\mathbmss D^{1/2} \mathbmss R \bmath b.
\end{equation}
This implies,
\begin{align}
Y(\bmath a,\mathbmss R,\bmath b) &= \int f(\bmath x) \psi\left(\left|\frac{\mathbmss R (\bmath x-\bmath b)}{\bmath a}\right|\right) d\bmath x \nonumber\\
 &= |a_1\ldots a_n| \int f(\bmath b + \mathbmss R^{\rm T} (\bmath a \odot \bmath t)) \psi(t) d\bmath t,
\label{ycwtss}
\end{align}
where $\odot$ stands for the coordinatewise vector multiply operation, and the division of
vector arguments $\frac{*}{*}$ is also understood coordinatewisely.

Finally, we introduce the so-called entirely isotropic case. Now we restrict the scale
matrix $\mathbmss A$ so that it is reduced to just a single scalar parameter: $\mathbmss D
= \kappa^2 \mathbmss I$, $\kappa=1/a$, and $\mathbmss R = \mathbmss I$. This yields
\begin{align}
Y(a, \bmath b) &= \int f(\bmath x) \psi\left(\frac{|\bmath x-\bmath b|}{a}\right) d\bmath x, \nonumber\\
\Upsilon(\kappa, \bmath c) &= \int f(\bmath x) \psi\left(|\kappa \bmath x+\bmath c|\right) d\bmath x, \quad \bmath c=-\kappa \bmath b.
\label{ycwti}
\end{align}
This case involves just $n+1$ degrees of freedom.

We add a note that $Y(a,\bmath b)$ is neither new nor special CWT, but merely just a result
of sampling a subspace from the definition domain of $Y(\bmath a, \mathbmss R, \bmath b)$.
This means that some information is just cropped by such a subsampling; in particular we
may loose some information about elongated elliptic patterns in $f(\bmath x)$. But
$Y(\bmath a, \mathbmss R, \bmath b)$ is a special case of the general definition
$Y(\mathbmss A, \bmath b)$ rather than argument sampling, because it is restricted to a
special subfamily of wavelets. In that case the associated reduction of degrees of freedom
appeared as a natural consequence of restricting the $\psi$. This did not trigger any
additional information loss, i.e. the use of~(\ref{ycwt}) instead of~(\ref{ycwtss}) does
not recover any new information if $\psi$ is radially symmetric.

\subsection{Inverting the multidimensional CWT}
Further on, we will deal mainly with the entirely isotropic case. The CWT is easily
invertible in such a case, and the inversion formula reads
\begin{align}
f(\bmath x) &= \frac{1}{C_{\psi\gamma}} \int\limits_0^\infty \frac{da}{a^{2n+1}} \int\limits_{\mathbb R^n} Y(a,\bmath b) \gamma\left(\frac{|\bmath x - \bmath b|}{a}\right) d\bmath b \nonumber\\
 &= \frac{1}{C_{\psi\gamma}} \int\limits_0^\infty \kappa^{n-1} d\kappa \int\limits_{\mathbb R^n} \Upsilon(\kappa,\bmath c) \gamma\left(|\kappa \bmath x + \bmath c|\right) d\bmath c, \nonumber\\
C_{\psi\gamma} &= \int\limits_0^\infty \hat\gamma(s)\hat\psi^*(s) \frac{ds}{s} = \frac{1}{S_n}\int\limits_{\mathbb R^n} \hat\gamma(\bmath s)\hat\psi^*(\bmath s) \frac{d\bmath s}{|\bmath s|^n},
\label{icwtiso}
\end{align}
where $\gamma(t)$ is a radially-symmetric `reconstruction kernel', which can differ from
$\psi$ in general, while $\hat\gamma$ and $\hat\psi$ are the corresponding Fourier images.
The constant $S_n$ is the surface area of a unit sphere in $\mathbb R^n$.

The formula~(\ref{icwtiso}) represents a minor generalization of the one given e.g. by
\citet{WangLu10}, but involving an arbitrary number of dimensions and arbitrary $\gamma$.
We could not find a reference for this more general case, so we provide a brief derivation
of~(\ref{icwtiso}) in the \ref{sec_ICWT}. We also provide there a discussion of some
subtleties concerning the correct understanding and practical use of this inversion formula
(they also appear important in the 1D case).

It seems that in more complicated cases the CWT can be inverted too, though the inversion
formula attains even more degrees of freedom. However, we do not consider such general
cases below.

In~(\ref{icwtiso}), the Fourier transforms $\hat\psi$ and $\hat\gamma$ are radially
symmetric functions themselves, and they may be expressed by a single-dimensional Hankel
transform:
\begin{equation}
\hat\psi(\bmath s) = \hat\psi(s) = \frac{(2\pi)^{\frac{n}{2}}}{s^{\frac{n}{2}-1}} \int\limits_0^\infty \psi(t) J_{\frac{n}{2}-1}(s t)\, t^{\frac{n}{2}} dt.
\label{FHn}
\end{equation}
For $n=2$, this simplifies to
\begin{equation}
\hat\psi(s) = 2\pi \int\limits_0^\infty \psi(t) J_0(s t)\, t\, dt.
\label{FH2}
\end{equation}

The integral for $C_{\psi\gamma}$ requires that $\hat\gamma(0) \hat\psi(0) = 0$, or the CWT
cannot be inverted using this reconstruction kernel. This condition represents the mutual
admissibility requirement for $\gamma$ and $\psi$. At least one of these functions should
integrate to zero, i.e. to have a vanishing first momentum (this appears entirely analogous
to the 1D case).

\subsection{Importance of the orthogonality of the wavelet basis}
How the wavelet transform~(\ref{ycwti}) and its inversion~(\ref{icwtiso}) are interrelated
and why we use different kernel functions $\psi$ and $\gamma$?

By combining~(\ref{ycwti}) and~(\ref{icwtiso}) and noting that it should hold for any input
function $f$, we may obtain a symbolic orthogonality relation:
\begin{align}
\int_0^\infty \frac{da}{a^{2n+1}} \int \gamma\left(\frac{|\bmath x - \bmath b|}{a}\right) \psi\left(\frac{|\bmath x'-\bmath b|}{a}\right) d\bmath b = C_{\psi\gamma} \delta(\bmath x'-\bmath x).
\end{align}
Of course, this equality is sensible only if $C_{\psi\gamma}$ is neither zero nor infinity.

In particular, if $\gamma\equiv \psi$ then we must require $C_{\psi\psi} < +\infty$, which
is also known as the admissibility condition on $\psi$. In this case the wavelet $\psi$
would form an orthogonal basis, viewed in the $(a,\bmath b)$ space. In fact, mathematically
it might be more convinient to use the same kernel $\psi$ in the both formulae, thus
dealing solely with orthogonal bases in the Hilbert space.

But in practice we usually apply the following transformation sequence:
\begin{equation}
f(\bmath x) \stackrel{\text{CWT}}{\longmapsto} Y(a,\bmath b) \stackrel{\text{noise thresholding}}{\longmapsto} Y_{\rm thr}(a,\bmath b) \stackrel{\text{CWT}^{-1}}{\longmapsto} \tilde f(\bmath x).
\label{ftf}
\end{equation}
Here, the noise thresholding stage mixes all the deal because $Y_{\rm thr}$ is no longer a
wavelet transform at all (notice that an arbitrary function of $a$ and $\bmath b$ is not
necessarily a valid wavelet transform of any $f(\bmath x)$). Therefore, the inversion
formula plays even more important role that it might seem: it should intelligently find
such an $\tilde f(\bmath x)$ which has a wavelet transform $\tilde Y$ close to $Y_{\rm
thr}$. Since~(\ref{icwtiso}) is a linear operator than it represents, basically, a
projection in the Hilbert space. It is not difficult to show that with $\gamma=\psi$ we
deal with an orthogonal projection minimizing the $L_2$ norm of $|\tilde Y-Y_{\rm thr}|$.
However, such a mimimizing is not necessarily useful in practice: in our case it might be
more useful to minimize the noise that appears in $\tilde f$ through the noisy $Y_{\rm
thr}$ (and coming, after all, from the input sample $\{x_i\}$). Then it might appear
reasonable to use a different kernel $\gamma$, and~(\ref{icwtiso}) then describes an
oblique projection rather than orthogonal one.

But playing with $\gamma$ in such a way we may achieve quite intriguing conclusions.
Formally, from~(\ref{icwtiso}) it follows that we can even adopt $\psi$ to be just a
Gaussian or other bell-shaped smoothing kernel, rather than a wavelet with vanishing first
momentum, and then use a wavelet-like shape for $\gamma$. Mathematically, the inversion
formula works if $\int\gamma =0$, but $\int\psi\neq 0$. Hence, even just a simple smoothing
transform can be easily inverted, if we assume an appropriate $\gamma$ in~(\ref{icwtiso}).
One may ask then: why we actually need wavelets and the CWT? Why not to use this more
simple setting with Gaussian $\psi$, thus trying to decompose $f(\bmath x)$ into a set of
Gaussian components of different scale and position?

However, such a setting does not appear practical exactly because the orthogonality of such
a Gaussian basis is lost. In this case it appears that different resolutions (i.e.,
different $a$-slices in $Y(a,\bmath b)$ function) become severely dependent. Basically, if
$\psi$ is merely a Gaussian then any large-scale slice of $Y(a,b)$ represents just a
smoothed version of the small-scale ones. Since all resolution layers become
interdependent, such an analysis would not be a truly multiresolution one. Or, in other
words, this means that our decomposition into elementary Gaussians would lack uniqueness:
we can easily approximate a wide Gaussian by a superposition of more narrow Gaussians,
disregarding the large-scale part. Thus, multiple resolutions are no longer informative in
this case.

In particular, applying in~(\ref{ftf}) a noise-filtering procedure to $Y(a,\bmath b)$ would
be rather meaningless in this case: masking out some piece of the $(a,\bmath b)$-space from
$Y(a,\bmath b)$ would not completely remove the associated pattern in the original function
$f(x)$. Moreover, after applying~(\ref{ycwti}) to the `denoised' $\tilde f(x)$, the noisy pattern
might even restore in $Y(a,\bmath b)$, as if it was not masked.

Therefore, even though we may use an `oblique' inversion formula with $\gamma\neq\psi$ for
some practical noise-control purpuses, for our goals it is necessary that the `orthogonal'
inversion formula with $\gamma=\psi$ at least exists, i.e. that $C_{\psi\psi}$ is finite.
This would mean that $\tilde f(x)$ can be represented (at least in theory) as a
superposition of the wavelets $\psi$ having different scale and position, although for
practical reasons we may build $\tilde f$ by using another kernel $\gamma$.

\subsection{Informal justification of the method via the smoothed Laplace operator}
Let $\varphi(\bmath t)$ is a smoothing kernel, or bell-like shape in $n$ dimensions. Put
\begin{equation}
\psi(\bmath t) = \Delta \varphi(\bmath t) = \sum_{i=1}^n \frac{\partial^2 \varphi}{\partial t_i^2}.
\end{equation}

Then, integrating~(\ref{ycwt2}) by parts,
\begin{equation}
Y(\mathbmss A, \bmath b) = \left(\mathbmss A \mathbmss A^{\rm T}\right) \cdot \int f''(\bmath x)\, \varphi\left(\mathbmss A^{-1} (\bmath x-\bmath b)\right) d\bmath x.
\end{equation}
Here $f''(\bmath x)$ is an $n\times n$ Hessian matrix, and dot $\cdot$ stands for the
scalar product of two matrices of equal size.

For a radially symmetric generating function,
\begin{equation}
\varphi(\bmath t) = \varphi(t) \implies \psi(\bmath t) = \psi(t) = \varphi''(t) + \frac{n-1}{t} \varphi'(t),
\end{equation}
we obtain
\begin{equation}
Y(\bmath a, \mathbmss R, \bmath b) = (\mathbmss R^{\rm T}\mathbmss D^{-1} \mathbmss R) \cdot \int f''(\bmath x) \varphi\left(\left|\frac{\mathbmss R (\bmath x-\bmath b)}{\bmath a}\right|\right) d\bmath x.
\end{equation}

For the entirely isotropic case,
\begin{align}
Y(a, \bmath b) &= a^2 \int \Delta f(\bmath x) \varphi\left(\frac{\left|\bmath x-\bmath b\right|}{a}\right) d\bmath x \nonumber\\
 &= a^{n+2} \int \Delta f(\bmath b + a \bmath t) \varphi(t) d\bmath t.
\end{align}
The last formula attains the most intuitive interpretation: the wavelet transform of $f(x)$
is equivalent to smoothing the Laplacian of $f(x)$ with the kernel $\varphi$. For example,
for small $a$ we have $Y(a, \bmath b) \simeq a^{n+2} \Delta f(\bmath b)$, if $\varphi$ is
normalized to have a unit integral.

The Laplacian $\Delta f$, in one of its equivalent definitions, determines how much the
function changes, in average, if we move by a small step in a random direction:
\begin{equation}
\Delta f(\bmath x_0) = \lim_{r\to 0} \frac{2n}{r^2} \left( \frac{1}{\sigma(S_r)}\int\limits_{S_r(\bmath x_0)} f(\bmath x) d\sigma(\bmath x) - f(\bmath x_0) \right).
\end{equation}
Here, $S_r(\bmath x_0)$ is a sphere of a small radis $r$ around $\bmath x_0$, and $d\sigma$
is its surface integration measure. Positive $\Delta f(\bmath x_0)$ means that $f(\bmath
x)$ grows in average, if we step by a small quantity from $\bmath x_0$. If simultaneously
$\Delta f$ is large, this likely indicates that we are near a local minimum of $f(\bmath
x)$, or at least inside a local concavity (considered relatively to the local tangent
plane). Negative $\Delta f$, if it is large in absolute value, likely indicates a local
maximum of $f(\bmath x)$ or at least a local convexity relative to the tangent plane.

Therefore, if we select $\psi$ to be the Laplacian of some smoothing kernel $\varphi$, the
resulting CWT would highlight possible local inhomogeneities: either p.d.f. leaks (or gaps)
or p.d.f. peaks (clumps of objects). It is important that all these inhomogeneities are
always considered relatively to some local large-scale background (tangent plane at $\bmath
x_0$). Therefore, it is not a big issue if a subtle statistical subfamily is polluted by
e.g. a large-scale underlying gradient of $f(\bmath x)$. Even if the local maximum
associated to such a subfamily is distorted by the gradient so heavily that it no longer
represents a local maximum in the strict meaning, it still appears as a local convexity
with the same $\Delta f(\bmath x_0)$. The wavelet $\psi$ is blind to linear components, so
it will `see' only that local convexity, filtering out any underlying background.

However, we note that Laplacian $\Delta f$ in $n>1$ dimensions is a less rigorous tool than
$f''$ for $n=1$. In two or more dimensions the single condition $\Delta f<0$ is not yet
enough to ensure that we are inside a local convexity, and hence to rigorously claim the
detection of a statistical clump. The condition $\Delta f<0$ only means that $f$ is
decreasing \emph{in average} whenever we step aside, but not necessarily decreasing when
stepping \emph{in either} direction. In other words, near a saddle point $\Delta f$ can be
negative too, but this does not yet imply the existence of a local convexity and hence of a
clump. However, we expect that in practice such cases can be easily identified on the CWT
visually, so investigation of just the Laplacian $\Delta f$ may be informative enough.

For the Gaussian $\varphi$ and $n=2$, we obtain the famous 2D MHAT wavelet:
\begin{equation}
\varphi(t) = e^{-\frac{t^2}{2}} \implies \psi(t) = (t^2-2) e^{-\frac{t^2}{2}}.
\label{wHerm}
\end{equation}
Based on what was said above, computing the isotropic CWT based on the 2D MHAT wavelet is
equivalent to Gaussian smoothing of the Laplacian $\Delta f(\bmath x)$.

\section{Statistical estimations of the CWT and patterns detection}
\label{sec_detect}
\subsection{Estimating the 2D wavelet transform from a random sample}
This part of the analysis method remains essentially the same as in the 1D case, but we
nonetheless need to reproduce its basic formulae. Assume that we aim to analyse the
distribution of a random quantity $x$, and this $x$ has the p.d.f. of $f(x)$, or `the
$x$-distribution'. As in the 1D case, the CWT~(\ref{ycwt}) can be viewed as the
mathematical expectation of another random quantity $y$:
\begin{equation}
y = \psi\left(\mathbmss A^{-1}(\bmath x- \bmath b)\right), \qquad Y(\mathbmss A, \bmath b) = \expect y.
\label{ydef}
\end{equation}
For each point in the space of the $(\mathbmss A, \bmath b)$ arguments, this quantity
generates the so-called `$y$-distribution'.

We use the same shorthand $\langle *\rangle$ for the sample-averaging operation:
\begin{equation}
\langle \phi \rangle := \frac{1}{N} \sum_{i=1}^N \phi_i \quad \text{or} \quad \langle \phi(x) \rangle := \frac{1}{N} \sum_{i=1}^N \phi(x_i).
\end{equation}

Then, the natural estimation of $Y$ is the sample mean of $y_i$:
\begin{equation}
\widetilde Y(\mathbmss A, \bmath b) = \langle y \rangle = \left\langle \psi\left(\mathbmss A^{-1}(\bmath x - \bmath b)\right) \right\rangle,
\label{wavest}
\end{equation}
or the sample wavelet transform (SWT). It is an unbiased and for large $N$ asymptotically
Gaussian estimate of $Y$. Its variance is
\begin{align}
D &:= \disp \widetilde Y = \frac{\disp y}{N} = \nonumber\\
&= \frac{1}{N} \left[ \int\limits_{-\infty}^{+\infty} f(\bmath x) \psi^2\left(\mathbmss A^{-1}(\bmath x - \bmath b)\right) d\bmath x - Y^2 \right],
\label{wavvar}
\end{align}
and it can be estimated by the classic unbiased variance estimate
\begin{equation}
\widetilde D = \frac{\langle y^2 \rangle - \langle y\rangle^2}{N-1}.
\label{wavvarest}
\end{equation}

Finally, our primary goodness-of-fit statistic represents the standardized version of
$\widetilde Y$. Let our null hypothesis be expressed as $Y(\mathbmss A, \bmath b) =
Y_0(\mathbmss A, \bmath b)$. Then,
\begin{equation}
z(\mathbmss A, \bmath b) = \frac{\widetilde Y(\mathbmss A, \bmath b) - Y_0(\mathbmss A, \bmath b)}{\sqrt{\widetilde D(\mathbmss A, \bmath b)}}
\label{zdef}
\end{equation}
represents our primary test statistic. As before, it is nothing more than just the Student
t-statistic for the sample $\{y_i\}$. Large values of $z$ indicate that at the given point
$(\mathbmss A, \bmath b)$, our null hypothesis failed to model the sample, meaning that
there is a statistically significant pattern with these scale and position.

\subsection{Computing 2D CWT significance levels in the approximation of Gaussian noise}
As in the 1D framework, we attempt to test the following hypotheses:
\begin{align}
H_0 &: Y(\mathbmss A, \bmath b) = Y_0(\mathbmss A, \bmath b) \text{ everywhere in } \mathcal D, \nonumber\\
H_A &: Y(\mathbmss A, \bmath b) \neq Y_0(\mathbmss A, \bmath b) \text{ somewhere in } \mathcal D,
\end{align}
where $\mathcal D$ is some predefined domain in the $(\mathbmss A, \bmath b)$ space, e.g.
the Gaussian domain.

The $H_0/Y_0$ is rejected if the following occurs:
\begin{equation}
z_{\max} > z_{\rm thr}, \quad z_{\max} = \max_{(\mathbmss A, \bmath b)\in \mathcal D} |z(a,b)|.
\label{zthrmax}
\end{equation}
A notice that $z_{\max}$ is a random quantity here (an extreme value of the random field
$z$). Basically, it is our test statistic. The critical significance level $z_{\rm thr}$ is
determined based on the distribution function of $z_{\max}$:
\begin{equation}
P_{\max}(z) = \Pr( z_{\max}<z | H_0 ).
\label{evd}
\end{equation}
The threshold level $z_{\rm thr}$ depends on the desired false alarm probability (FAP) as
$\FAP = 1-P_{\max}(z_{\rm thr})$. Or, alternatively, we can derive the FAP based on the
observed $z_{\max}$, and then the associated significance \citep{Baluev18b}.

To accomplish this testing procedure, we must derive the extreme value distribution (EVD),
$P_{\rm max}(z)$, for the random field $z(\mathbmss A, \bmath b)$. As in the 1D case, we
derive this distribution analytically, assuming that $z(\mathbmss A, \bmath b)$ is standard
Gaussian inside $\mathcal D$, and applying the so-called Rice method \citep{AzaisDelmas02}.
Here we again rely on the intermediary theory from \citep{Baluev13b}.

We do not discuss the details of this theory, but we note that the Rice method allows one
to derive an efficient analytic approximation to $P_{\max}(z)$, for a Gaussian or even
non-Gaussian random process or field. The Rice approximation works asymptotically for
$z\to\infty$, i.e. for the more practical small $\FAP$ values. Simultaneously, such an
approximation appears to limit the $\FAP$ from the upper side. The latter property is very
important, because it means that this method never increases the fraction of false alarms:
even if the Rice approximation appeared inaccurate, the actual $\FAP$ value may be only
smaller, so any features detected in the CWT would remain significant. Given this, we
recognize that because of high task complexity the formulae presented below neglect some
secondary terms of the Rice approximation (in particular, those related to the extrema
attained on the boundary of $\mathcal D$). So the upper-limit property might appear
formally broken from the mathematical point of view. Also, it might appear broken if we
take into account the non-Gaussian deviations of $z$. However, simulation tests (see
Sect.~\ref{sec_tests}) revealed that in practically reasonable ranges this property remains
satisfied, confirming that the neglected terms keep small.

In fact, all necessary formulae can be easily derived from the 1D case with only minor
changes to the formulae, replacing a two-dimensional Gaussian random field by the
three-dimensional one. We consider here only the case of entirely isotropic CWT, which
implies a $3$-dimensional random field $z(a,\bmath b)$.

The one-sided false alarm probability is then given by
\begin{equation}
\FAP^+(z) \sim A_0 \left(\frac{z^2}{2}+1\right) e^{-\frac{z^2}{2}} + A_1 e^{-\frac{z^2}{2}},
\label{fap1}
\end{equation}
This is a special case of formula~(20) from \citep{Baluev13b}. In this work we keep only
the primary term in~(\ref{fap1}), assuming $z$ is large. Then we obtain:
\begin{equation}
\FAP^+(z) \sim W_{00} z^2 e^{-\frac{z^2}{2}},
\label{FAPgauss}
\end{equation}
where $W_{00}=A_0/2$ is given by
\begin{equation}
W_{00} = \frac{1}{4\pi^2} \int\limits_{\mathcal D} \sqrt{\det \mathbmss G(a,\bmath b)}\, da d\bmath b, \quad \mathbmss G = \var z',
\label{W00}
\end{equation}

The matrix $\mathbmss G$ in the last formula represents the covariance matrix of the 3-dim
gradient $z'$. It can be asymptotically approximated as in the 1D case:
\begin{equation}
\lim_{N\to\infty} G_{ij} = \frac{\cov(y'_i,y'_j)}{\disp y} - \frac{\cov(y,y'_i)\cov(y,y'_j)}{(\disp y)^2},
\label{zgvar}
\end{equation}
where the index denotes the differentiation variable, either $a$ or one of the $\bmath b$
components.

In practical computations, we do not know the covariances appearing in~(\ref{zgvar}),
because they depend on the unknown p.d.f. $f(\bmath x)$ which we aim to analyse. However,
we may replace them by the corresponding sample estimates:
\begin{align}
\cov(y'_i,y'_j) &\simeq \langle y'_i y'_j \rangle - \langle y'_i \rangle \langle y'_j \rangle, \nonumber\\
\cov(y,y'_i) &\simeq \langle y y'_i \rangle - \langle y \rangle \langle y'_i \rangle, \nonumber\\
\disp y &\simeq \langle y^2 \rangle - \langle y \rangle^2.
\label{covest}
\end{align}

The formulae~(\ref{zgvar}) and~(\ref{covest}) allow us to compute an estimate
$\widetilde{\mathbmss G}$ that we can substitute in~(\ref{W00}) in place of $\mathbmss G$
and then integrate it numerically. This substitution infers a relative error of $\sim
1/\sqrt N$ in the result.

Finally, to treat the two-sided estimation $\FAP$ we must double~(\ref{FAPgauss}):
\begin{equation}
\FAP(z) \sim 2 W_{00} z^2 e^{-\frac{z^2}{2}}.
\label{dsFAPgauss}
\end{equation}

\subsection{Determining the Gaussian domain in the 2D CWT}
\label{sec_gauss}
As noticed above, individual values of the statistic $z$ are asymptotically standard normal
for $N\to\infty$, if the null hypothesis $Y_0$ is true. Hence, $z$ is an asymptotically
Gaussian random field. However, for a finite $N$ there are always points in the $(a,\bmath
b)$ space, where the normality is broken too much. In particular, the normality is always
violated at small enough $a$, where the wavelets become so narrow-localized that we have
just a few or even one term dominating in the sum~(\ref{wavest}). In general, we can
determine some limited `normality domain' $\mathcal D$, depending on $N$, such that inside
$\mathcal D$ the random field $z$ is close to being Gaussian, but essentially non-Gaussian
outside. To determine $\mathcal D$ we need to define some reasonable normality indicators.

As in the 1D case, the first rough indicator of the normality is the effective number of
terms contributing to the sum~(\ref{wavest}). It can be roughly estimated by:
\begin{equation}
n(\mathbmss A, \bmath b) = \left[ \sum_{i=1}^N \varphi\left(\mathbmss A^{-1}(\bmath x- \bmath b)\right) \right] \left/ \int\limits_{-\infty}^{+\infty} \varphi(x) dx \right. .
\label{num}
\end{equation}
Since $\varphi$ has approximately the same localization as $\psi$, the number of terms
dominating in~(\ref{num}) is approximately the same as in~(\ref{wavest}), but now all these
terms are positive and do not cancel.

We also tried to follow a more rigorous method to construct the normality indicator, based
on the Edgeworth-type decomposition of the $\FAP$. This method is detailed in
\citep{Baluev18a}. It remains almost unchanged here, but now the dimension of the gradient
$z'$ is increased from $2$ to $3$, while the dimension of the Hessian matrix $z''$ becomes
$3\times 3$ with $6$ independent elements. Therefore, the total dimension of the random
vector $(z,z',z'')$, which plays the central role in this method, is increased from $6$ to
$10$. Unfortunately this apparently moderate increase of dimensionality triggered a huge
increase of algebraic complexity of the task, so we could not replicate the derivation even
with the help of computer algebra. We only give here a reduced result, which appeared to
convolve to a simple final form regardless of a complicated derivation. The asymptotic
distribution of the one-sided false alarm probability appears to have the following
double-series shape:
\begin{align}
\FAP^\pm(z) &\sim z^2 e^{-z^2} \times \left[ \vphantom{\frac{1}{N}} \left(W_{00} + \mathcal O(z^{-2}) + \ldots \right) \pm \right. \nonumber\\
&\quad \pm \frac{z^3}{\sqrt N}\left( W_{1,-3} + \mathcal O(z^{-2}) + \ldots \right) + \nonumber\\
&\quad \left. + \mathcal O\left(\frac{z^6}{N}\right) + \ldots \right], \nonumber\\
W_{ij} &= \frac{1}{(2\pi)^2} \int\limits_{\mathcal D} q_{ij}(a,\bmath b) \sqrt{\det \mathbmss G(a,\bmath b)}\, da d\bmath b.
\label{FAPngauss}
\end{align}
Basically it remained the same as in the 1D case, except for the change $z$ to $z^2$ in the
very first factor. The first part of the sum that does not involve $N$, represents the
Gaussian part, while the rest is a non-Gaussian perturbation. Here we left just one leading
term in each series decomposition, so we only have the primary 'Gaussian' coefficient
$W_{00}$, which has $q_{00}=1$, and the first-order non-Gaussian correction with
$W_{1,-3}$, defined via the quantity $q_{1,-3}(a,\bmath b)$. The latter appears to have the
same form as in the 1D case:
\begin{equation}
q_{1,-3} = -\frac{1}{3} \As y,
\label{qij}
\end{equation}
where $\As y$ is the skewness of the $y$-distribution. In practice, $\As y$ in~(\ref{qij})
can be estimated simultaneously with the SWT~(\ref{wavest}) and matrix $\mathbmss G$
in~(\ref{zgvar}), using the sample skewness of $y_i$.

As in the 1D case, we do not recommend the use of~(\ref{FAPngauss}) directly because of its
likely poor convergence, and because we actually have just one correction term there.
Instead, we may use this formula to remove the points $(a,\bmath b)$ that contribute too
much in the cumulative non-Gaussian term.

Namely, let us rewrite~(\ref{FAPngauss}) as
\begin{align}
\FAP^\pm(z) = \int\limits_{\mathcal D} \mu(z, a,\bmath b) da d\bmath b, \quad \mu = \mu_\mathrm{gauss} + \mu_\mathrm{nongauss}, \nonumber\\
\mu_\mathrm{gauss} \simeq \frac{z^2 e^{-z^2}}{(2\pi)^2}\sqrt{\det \mathbmss G}, \quad
\frac{\mu_\mathrm{nongauss}}{\mu_\mathrm{gauss}} \sim \pm \frac{z^3}{\sqrt N} q_{1,-3}.
\end{align}
Now, if we restrict the domain $\mathcal D$ to only such points where
\begin{equation}
\left| \frac{\mu_\mathrm{nongauss}}{\mu_\mathrm{gauss}} \right| \leq \varepsilon
\end{equation}
then we will have $|\FAP_\mathrm{nongauss}| \leq \varepsilon \FAP_\mathrm{gauss}$, i.e. the
relative magnitude of the non-Gaussian error in the $\FAP$ is bounded by the same threshold
$\varepsilon$.

Therefore, the necessary normality criterion can be formulated as follows:
\begin{equation}
 \frac{z_*^3}{\sqrt N}\, |q_{1,-3}(a, \bmath b)| < \varepsilon,
\label{nrmtest}
\end{equation}
which represents a reduced version of a similar criterion for the 1D case.

The meaning of this $z_*$ in~(\ref{nrmtest}) is the maximum $z$-level, for which the
normality will be guaranteedly preserved with the desired accuracy.\footnote{It is typical
that relative magnitude of a non-Gaussian error in a density function grows infinitely in
the tails. Therefore, we have to limit $z$ by some $z_*$, below which we expect to
constrain the non-Gassianity.} This $z_*$ may be derived from the desired smallest
considerable FAP level $\FAP_*$ as $\FAP_* = \FAP_\mathrm{gauss}(z_*)$. It therefore
depends on many input paraameters (including $W_{00}$), but nonetheless it appears
relatively insensitive to all them, thanks to roughly logarithmic dependence with respect
to $\FAP_*$. In the 1D case we recommended to put $z_*\sim 3$ in practice, while in the 2D
case a bit more safe value $z_* \sim 4-6$ may be used.

Another parameter $\varepsilon$ sets the desired limit on the relative magnitude of
possible non-Gaussian corrections to the FAP at $z=z_*$. We usually set
$\varepsilon^2=0.1$. Using these parameters, the normality domain in the 1D case appeared
generally close to the one obtained from the simplistic criterion $n(a,\bmath b)\geq
n_{\min}$ with $n_{\min}=10$.

Since we neglected in~(\ref{nrmtest}) several additional terms, this criterion can be
simplified to the following shape:
\begin{equation}
 | \As y | < \frac{3\varepsilon}{z_*^3} \sqrt N = \frac{\sqrt N}{\const}, \quad \const\sim 100-300.
\label{nrmtest2}
\end{equation}
Here, $\As y$ can be estimated from $y_i$ by the corresponding sample momentum.

As in the 1D case, the skewness $\As y$ plays a predominant role here. It determines the
non-Gaussian error of the largest order. Therefore, we can suppress the non-Gaussianity and
simultaneously significantly expand the domain $\mathcal D$ if we select such a wavelet
$\psi$ that implies a reduced $\As y$.

Here we need to discuss a yet another subtle statistical issue missed in the 1D paper.
Given the formulae above, the domain $\mathcal D$ is determined based on the input sample
$\{x_i\}$, and as such it may not be considered as an a priori \emph{predefined} domain. As
a general rule, it might be a bad and potentially dangerous practice to replace an \emph{a
priori} information required by a statistical test with an \emph{a posteriori} estimate
instead. Such a substitution may lead to unexpected statistical biases sometimes, since it
violates the conditions of the original statistical test. However, the domain determined
by~(\ref{nrmtest}) is a sample-based \emph{approximation} to some `theoretic' normality
domain $\mathcal D$ that does not depend on individual $x_i$ in the input sample. The
relative error of this approximation should remain small, corresponding to the typical
statistical error $\sim 1/\sqrt N$. Therefore, the effect of random $x_i$ on the derived
$\mathcal D$ (and hence~--- on the coefficient $W_{00}$) is in this case negligible, and
this $\mathcal D$ mainly depends on the actual p.d.f. $f(\bmath x)$, on the wavelet $\psi$,
and on the sample size $N$.

\section{Deriving optimal minimum-noise wavelets}
\label{sec_opt}
\subsection{2D wavelets normalization}
The generating function must satisfy the following restrictions:
\begin{eqnarray}
\varphi(t)>0 \; {\rm everywhere}, \quad \varphi(t)=\varphi(-t), \nonumber\\
\varphi'(t)>0 \; {\rm for} \; t<0, \quad \varphi'(t)<0 \; {\rm for} \; t>0.
\label{gfc}
\end{eqnarray}
These conditions define some domain $\Pi$ in the functional space. Notice that the
condition $\varphi(t)>0$ is a bit excessive here, because if $\varphi'(t)<0$ is satisfied
for $t>0$ then it is enough to require that $\varphi(t)>0$ for just $t\to\infty$. The
positiveness of $\varphi$ for all other $t$ then follows automatically.

Similarly to the 1D case, to legally compare different wavelets between each other, we
formulate the following normalization constraints:
\begin{equation}
\int\limits_0^{+\infty} \varphi(t)\, t dt = \frac{1}{2\pi}, \qquad \int\limits_0^{+\infty} \psi^2(t)\, t dt = \frac{1}{2\pi}.
\label{wavnrm}
\end{equation}
The first of the equalities~(\ref{wavnrm}) naturally requires $\varphi$ to be a normalized
smoothing kernel, so that the CWT always has the same magnitude and dimension as $\Delta
f(\bmath x)$. This sets a `standard' signal normalization in the CWT. As in the 1D case,
the second equation of~(\ref{wavnrm}) fixes a `standard' noise normalization in the SWT,
namely it fixes the magnitude of $D(a,\bmath b)$ given a small scale $a$.

An arbitrary wavelet $\psi=\Delta\varphi$ can be forced to fulfill~(\ref{wavnrm}) by the
following affine transformation:
\begin{align}
\varphi_{\rm nrm}(t) &= K \varphi(kt), \nonumber\\
\psi_{\rm nrm}(t) &= Kk^2 \psi(kt), \nonumber\\
K &= (2\pi)^{-\frac{2}{3}} \left(\int\limits_0^{+\infty} \varphi\, t dt\right)^{-\frac{1}{3}} \left(\int\limits_0^{+\infty} \psi^2 t dt \right)^{-\frac{1}{3}}, \nonumber\\
k &= (2\pi)^{\frac{1}{6}} \left(\int\limits_0^{+\infty} \varphi\, t dt \right)^{\frac{1}{3}} \left(\int\limits_0^{+\infty} \psi^2 t dt \right)^{-\frac{1}{6}}.
\label{evennrm}
\end{align}

In particaluar, for the 2D MHAT wavelet we obtain:
\begin{equation}
\text{2DMHAT} : \quad K=0.2936839, \; k=1.358407.
\end{equation}

\subsection{Finding optimal 2D wavelet}
Analogously to the 1D case, for small $a$ the skewness of the $y$-distribution becomes approximated by
\begin{equation}
\As y \approx \frac{\mathcal R}{\sqrt{a^2 f(\bmath b)}},
\end{equation}
where
\begin{equation}
\mathcal R = 2\pi \int\limits_0^{+\infty} \psi_{\rm nrm}^3(t)\, t dt.
\end{equation}

Therefore, $\As y$ is drastically reduced if the following condition is satisfied:
\begin{equation}
\int\limits_0^{+\infty} \psi^3(t)\, t dt = 0.
\label{psi3}
\end{equation}
This is not fulfilled by the 2D MHAT which has $\mathcal R \approx -0.64$. In the 1D case,
we had the analogous proportionality coefficient $\mathcal R = -0.49$, so matters are
getting worse in 2D.

If the normalization~(\ref{wavnrm}) is satisfied, the objective functional to minimize for
optimality becomes:
\begin{equation}
a^6 \det \mathbmss G \approx T(\psi) = \left(2\pi \int\limits_0^{+\infty} \psi'^2 t^3 dt - 1 \right) \left(\pi \int\limits_0^{+\infty} \psi'^2 t dt \right)^2.
\label{detGss}
\end{equation}
This approximation is obtained analogously to the 1D case, by assuming that $a$ is small
and $f(\bmath x)$ is constant within the wavelet localization domain.

The minimization of~(\ref{detGss}) allows to minimize the coefficient $W_{00}$~(\ref{W00})
in the FAP representation~(\ref{FAPgauss}), assuming that the domain $\mathcal D$ is
fixed.

An arbitrary input wavelet must be first transformed to the appropriate scales
using~(\ref{evennrm}), and then the resulting $\psi_{\rm nrm}$ should substituted to
$T(\psi)$. This yields us the following objective functional:
\begin{align}
T_{\rm nrm} &= \frac{(2\pi)^{\frac{2}{3}}}{4} \left(\int\limits_0^{+\infty} \psi'^2 t^3 dt - \int\limits_0^{+\infty} \psi^2 t dt \right) \left(\int\limits_0^{+\infty} \varphi\, t dt\right)^{\frac{4}{3}} \times \nonumber\\
&\quad \times \left(\int\limits_0^{+\infty} \psi^2 t dt \right)^{-\frac{11}{3}} \left(\int\limits_0^{+\infty} \psi'^2 t dt \right)^2.
\label{Tnrmeven}
\end{align}

From~(\ref{W00}) and~(\ref{detGss}), the coefficient $W_{00}$ becomes
\begin{equation}
W_{00} \approx \frac{1}{4\pi^2} \int\limits_{\mathcal D} \sqrt{T_{\rm nrm}}\, \frac{da d\bmath b}{a^3},
\label{W00Tnrm}
\end{equation}
so in practice the values of $\mathcal W = \sqrt{T_{\rm nrm}}/(2\pi)^2$ might have perhaps
a more intuitive interpretation than $T_{\rm nrm}$ itself. Namely, we have $W_{00} \simeq
\mathcal W \Vol(\mathcal D)$, where $\Vol(\mathcal D)$ is the total volume of the domain
$\mathcal D$, considered in the space of $(\log a, \bmath c)$, but without the large-scale
part $a\gtrsim \disp x$, where the approximation~(\ref{detGss}) does not work (as numeric
tests confirmed, this large-scale portion in $W_{00}$ is usually negligible).

The scale-invariant quantity $\mathcal W$ describes how noisy a particualar wavelet is. For
example, the 2D MHAT wavelet has $\mathcal W = 0.12$, which means very low noise. However,
this low noise level is achieved by the cost of a large noise skewness, which is not
acceptable in this work.

Since all integrals above involve the integration measure $t dt = d(t^2/2)$, it is now
convenient to use $u=t^2/2$ as an auxiliary proxy variable:
\begin{align}
\varphi(t) = \tilde\varphi\left(\frac{t^2}{2}\right), \quad
\psi(t) = \tilde\psi\left(\frac{t^2}{2}\right), \nonumber\\
\tilde\psi(u) = 2u \tilde\varphi''(u) + 2 \tilde\varphi'(u).
\end{align}

Now, the parametric wavelet model from \citep{Baluev18a} can be rewritten as:
\begin{align}
\tilde\varphi(u) &= P(u) e^{-u}, &P(u) &= p_0 + p_1 u + \ldots + p_m u^m, \nonumber\\
\tilde\varphi'(u) &= P_1(u) e^{-u}, &P_1(u) &= P'(u) - P(u), \nonumber\\
\tilde\varphi''(u) &= P_2(u) e^{-u}, &P_2(u) &= P_1'(u) - P_1(u), \nonumber\\
\tilde\psi(u) &= P_\Delta(u) e^{-u} &P_\Delta(u) &= 2 u P_2(u) + 2 P_1(u).
\label{wmp}
\end{align}
Substituting this model to $T_{\rm nrm}$ yields an algebraic objective function in $p_k$,
and another algebraic equality constrant follows from~(\ref{psi3}). We need to find such
$p_k$ that minimize the objective conditionally to the constraint. This optimization task
can be solved using the Largange multipliers approach.

However, the polynomial $P(u)$ cannot be arbitrary. Since $\varphi(t)$ should belong to the
domain $\Pi$ defined in~(\ref{gfc}), the following equivalent restrictions must be
satisfied:
\begin{equation}
P(u)>0 \; {\rm for} \; u\to +\infty, \quad P_1(u)<0 \; {\rm for} \; u>0.
\end{equation}
Assuming without loss of generality that $p_m = +1$, this guarantees $P(u)>0$ and
$P_1(u)<0$ for large $u$. Then the constraints can be reformulated equivalently as just
\begin{equation}
P_1(u) \; \text{has no real roots for} \; u>0.
\label{gfcP}
\end{equation}
We must discard all local optima outside of the domain $\Pi$, i.e. all solutions $\{p_k\}$
for which $P_1(u)$ has positive real roots.

But it is also necessary to verify possible boundary optima. The boundary of $\Pi$ can be
mathematically determined by violating~(\ref{gfcP}) in a way to obtain a transitional
state. Condition `the polynomial has no positive real roots' can be violated so in two
ways. The first one is to require that $u=0$ is a root, equivalent to setting the free term
of $P_1$ to zero, or setting $p_1=p_0$. The second way is a bit more tricky. Since we need
to trigger just a boundary state, we may request that `the polynomial has double or
multiple roots', which includes all occurences whenever a new pair of roots is about to
appear (possibly, but not necessarily, at $u>0$). This is equivalent to equating the $P_1$
discriminant $D(p_1,\ldots p_m)$ to zero, therefore defining an extra algebraic equality
constraint in terms of $\{p_k\}$. Not all occurences with $D=0$ necessarily indicate the
boundary of $\Pi$, because e.g. the emerging double root $u$ may be negative, which does
not imply any violation in~(\ref{gfcP}), or we could already had have other roots $u>0$ and
thus are already well outside of the domain $\Pi$. But regardless of that, the nonlinear
subspace $D=0$ necessarily includes the boundary of $\Pi$, so we can just seek all optima
in this subspace, then post-check whether they are actually located at the boundary, and if
yes whether they supply a smaller minimum of $T_{\rm nrm}$ than any local minimum already
found in the domain interior. This will complete the comprehensive search of optimal
solutions in $\Pi$.

We performed a comprehensive search of wavelets represented by the model~(\ref{wmp}) for
$m=1$ and $m=2$, but we surprisingly found none at least satisfying~(\ref{psi3})
and~(\ref{gfc}) simultaneously. The cases $m=3$ and $m=4$ are too difficult to be processed
analytically, even with computer algebra, but extensive numeric search also did not reveal
any wavelet at least satisfying the constraints, so we could not even start the search for
optimality. This contrasts with the 1D case from \citep{Baluev18a}, where we did found the
necessary optimal wavelet already for $m=2$.

In order to acquire a more deep understanding of the issue, we considered the model of
another type:
\begin{equation}
\varphi(t) = e^{-|t|^p}, \quad p > 0.
\end{equation}
Here, parameter $p$ regulates the wings of the generating function: larger $p$, more
abruptly they decrease to zero. We found that the condition~(\ref{psi3}) is then satisfied
for $p=6.34$. We additionally calculated that in the 1D case this condition is satisfied
for a considerably smaller degree $p=3.26$. That is, the 2D case likely requires the tails
of $\varphi$ to be almost twice as abrupt as in the 1D case. Consequently, the kernel part
of $\varphi$ must be relatively wider in 2D.

Sadly, unskewed models of this type are unpleasantly more noisy, e.g. we had $\mathcal
W=2.17$ for the $p=6.34$ 2D solution. To a certain extent, an increase of $\mathcal W$
seems to be a necessary sacrifice paid for the vanishing noise skewness. Note that
non-Gaussian noise is poorly predictable and difficult to model, even if it has a smaller
overall magnitude, so we put more priority on reducing its deviation from the normality.

Since in the 1D case we could find a suitable wavelet using the model with square-exponent
tails, $\exp(-t^2)$, in the 2D case we likely need a quart-exponent law, $\exp(-t^4)$. This
means that $\exp(-u)$ in~(\ref{wmp}) should be replaced by something like $\exp(-u^2)$. The
search using such parametric model for $m=2$ succesfully resulted in the following wavelet:
\begin{align}
\tilde \varphi(u) &= (2.73050 - 1.00854 u + u^2) e^{-\frac{u^2}{2}}, \nonumber\\
\tilde \psi(u) &= ( -2.01708 - 2.92199 u + 8.06833 u^2 - \nonumber\\
&\quad - 6.53900 u^3 - 2.01708 u^4 + 2 u^5) e^{-\frac{u^2}{2}}, \nonumber\\
&K=0.0796646, \; k=1.35480.
\label{2DOPT1}
\end{align}
It has $\mathcal W=1.07$. Below we will refer it as `2DOPT1' wavelet.

However, functions involving $\exp(-t^4)$ did not appear to have an easy analytic form of
the Fourier transform. So we continued to seek yet another wavelet with a more practical
functional form.

We should consider symmetric $\varphi$ that have depressed tails, i.e. zero skewness and
decreased kurtosis (speaking as if $\varphi$ was a p.d.f.). Among the general class of
Pearson distributions, only the type II family satisfies this constraint. This is basically
the family of symmetric beta distribution, or `powers of semicircle'. If $\varphi$ is such
a symmetric beta distribution, put
\begin{equation}
\tilde \varphi_\nu(u) = (1-u)^\nu, \quad u\leq 1, \quad \nu \geq 0.
\label{psc}
\end{equation}
Then from~(\ref{FH2}) we derive the Fourier transform
\begin{align}
\hat\varphi_\nu(s) &= 2\pi \int\limits_0^{\sqrt{2}} \left(1-\frac{t^2}{2}\right)^\nu J_0(s t)\, t\, dt = \nonumber\\
 &= 2\pi \Gamma(\nu+1) \left(\frac{\sqrt{2}}{s}\right)^{\nu+1} J_{\nu+1}\left(s\sqrt{2}\right),
\end{align}
which behaves closely to the usual bell-like shape (with only negligible wavy oscillations
in the tails). Note that the last equality was suggested by the machine computer algebra,
so we do not provide its derivation. For $\nu\to\infty$ we come to the Gaussian $\varphi$
and 2D MHAT $\psi$, which is too much skewed in terms of~(\ref{psi3}). But for $\nu=2$ we
have such $\psi$ that~(\ref{psi3}) is remarkably satisfied. Unfortunately, this simple
special case is still unsuitable for us, because such $\psi$ has jumps at $t = \pm 1$,
hence the derivative $\psi'$ is singular and the objective $T_{\rm nrm}$ becomes infinite.

Therefore, we may seek the suitable wavelet in the following parametric form, constructed
analogously to the Gaussian case:
\begin{align}
\tilde\varphi(u) &= (1-u)^\nu P(u), \nonumber\\ &P(u) = p_0 + p_1 u + \ldots + p_m u^m, \nonumber\\
\tilde\varphi'(u) &= (1-u)^{\nu-1} P_1(u), \nonumber\\ &P_1(u) = (1-u) P'(u) - \nu P(u), \nonumber\\
\tilde\varphi''(u) &= (1-u)^{\nu-2} P_2(u), \nonumber\\ &P_2(u) = (1-u) P_1'(u) - (\nu-1) P_1(u), \nonumber\\
\tilde\psi(u) &= (1-u)^{\nu-2} P_\Delta(u), \nonumber\\ &P_\Delta(u) = 2 u P_2(u) + 2 (1-u) P_1(u),
\label{wbeta}
\end{align}
where $u\in [0,1]$. Now we optimize against $p_k$ and $\nu$ simultaneously ($\nu$ is not
necessarily an integer). Note that although the definition~(\ref{psc}) required $n\geq 0$, the
quantity $T_{\rm nrm}$ is defined and finite only if $\nu > \frac{5}{2}$.

Using this parametric form, we found the first optimal solution already for $m=1$: $\nu = 2.98010$
with $p_0=0.397808$ and $p_1=1$. This optimum is located very close to the boundary of
$\Pi$, but still in the interior. However, it corresponds to $\mathcal W=1.83$ which is
still too large for us, e.g. the wavelet~(\ref{2DOPT1}) is less noisy.

The next optimal solution was obtained for $m=2$. It corresponds to the following wavelet:
\begin{align}
\tilde \varphi(u) &= (0.141826 + 0.330025 u + u^2) (1-u)^\nu = \nonumber\\
 &= 1.471851 (1-u)^\nu - 2.330025 (1-u)^{\nu+1} + \nonumber\\
 &\quad + (1-u)^{\nu+2}, \nonumber\\
\tilde \psi(u) &= (-0.584764 + 3.452872 u - 40.719896 u^2 +\nonumber\\
&\quad + 81.626277 u^3) (1-u)^{\nu-2} = \nonumber\\
 &= 43.774488 (1-u)^{\nu-2} - 166.891909 (1-u)^{\nu-1} + \nonumber\\
 &\quad + 204.158934 (1-u)^\nu - 81.626277 (1-u)^{\nu+1}, \nonumber\\
&\nu=4.388516, \quad u\leq 1, \nonumber\\
&K=1.54811, \; k=0.652500.
\label{2DOPT2}
\end{align}
Below we will refer it as `2DOPT2' wavelet. It has $\mathcal W=1.05$, which is almost the
same as for 2DOPT1.

\begin{figure*}[!t]
\includegraphics[width=\textwidth]{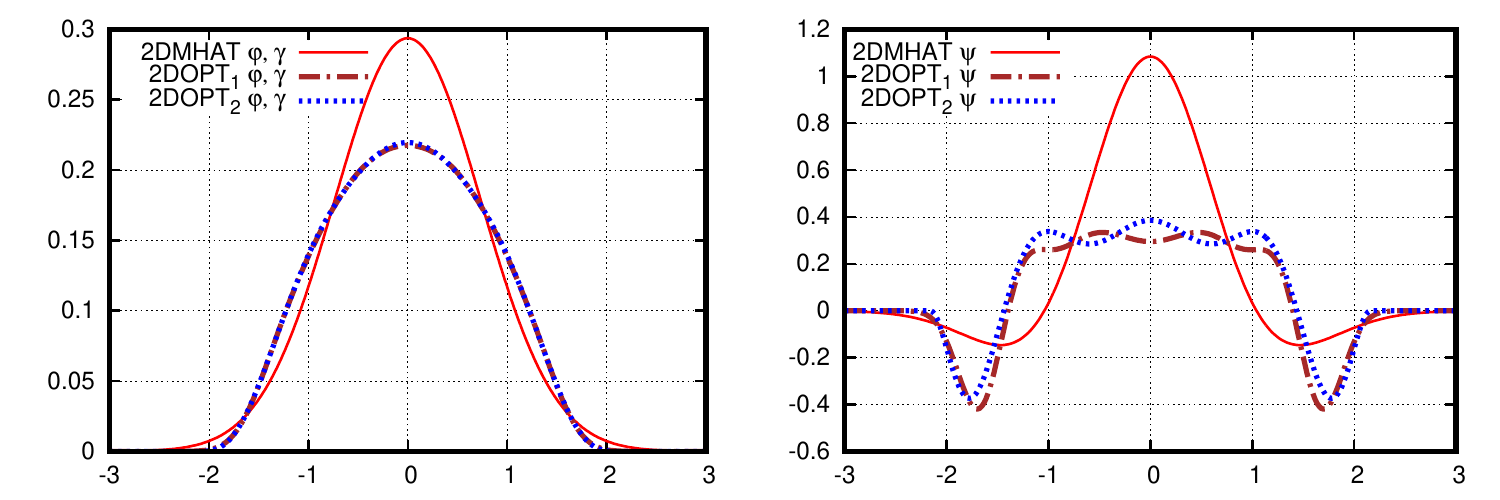}
\caption{Radial functions for two optimal 2D wavelets ($\psi$) and for their generators
($\varphi$), compared with $\psi$ and $\varphi$ of the classic 2D MHAT wavelet. In the 2D
case, the optimal generating function $\gamma$ coincides with $-\varphi$. Everything shown
after the normalization~(\ref{wavnrm}).}
\label{fig_wavs}
\end{figure*}

The graphical shape of~(\ref{2DOPT2}) is also almost identical to~(\ref{2DOPT1}), so
numerically they appear to be practically the same wavelets. However, the functional shape
of~(\ref{2DOPT2}) is more analytically tractable in the Fourier space.

Since we obtained almost the same shape using two different functional models, we believe
that further improvement of $T_{\rm nrm}$ is not very likely, if we do not violate some
natural restrictions of smoothness and good convergence, or the normality
condition~(\ref{psi3}). It looks as if $\mathcal W\approx 1$ was a natural lower limit on
possible noise in the 2D SWT.

\subsection{Finding optimal reconstruction kernel}
Entirely analogously to the 1D case, we may obtain that the variance of the estimated
density $\tilde f(x)$, obtained by applying the inversion formula~(\ref{icwt}) in some
finite `signal domain' $\mathcal S$, is
\begin{equation}
N \disp \tilde f(\bmath x) \simeq f(\bmath x) \int\limits_{\mathbb R^n} R_{\mathcal S}^2(\bmath x, \bmath x') d\bmath x',
\end{equation}
where
\begin{equation}
R_{\mathcal S}(\bmath x,\bmath x') = \frac{1}{C_{\psi\gamma}} \iint\limits_{\mathcal S} \psi\left(\frac{\bmath x - \bmath b}{a}\right) \gamma\left(\frac{\bmath x - \bmath b}{a}\right) \frac{da d\bmath b}{a^{2n+1}}.
\label{Rfunc}
\end{equation}
The cumulative relative variance characteristic then becomes:
\begin{equation}
N \int\limits_{\mathbb R^n} \frac{\disp \tilde f(\bmath x)}{f(\bmath x)} d\bmath x \simeq
\iint\limits_{\mathbb R^n \times \mathbb R^n} R_{\mathcal S}^2(\bmath x, \bmath x') d\bmath x d\bmath x'.
\label{fen}
\end{equation}

We therefore should minimize the $L_2$ norm $\|R_{\mathcal S}\|$ with respect to $\gamma$
to obtain the least noisy p.d.f. reconstruction:
\begin{equation}
\| R_{\mathcal S} \|^2 = \iint\limits_{\mathbb R^n \times \mathbb R^n} R_{\mathcal S}^2(\bmath x,\bmath x') d\bmath x d\bmath x' \longmapsto \min_\gamma
\end{equation}
Note that the effect of the domain $\mathcal S$ is crucial here, because e.g. if $\mathcal
S$ spans the entire $(a,\bmath b)$ space then $R_\infty = \delta(\bmath x - \bmath x')$
thanks to the wavelet orthogonality, and the variance $\disp \tilde f$ gets therefore
infinite. This reflects the known fact that p.d.f. estimation based on a finite sample is
an ill-posed task: we must cut off at least small scales in~(\ref{icwtiso}) to obtain a
meaningful reconstruction (or we just `reconstruct' the input sample as a comb of delta
functions, rather than any smooth p.d.f.).

As in the 1D case, we consider the optimality in the local sense, i.e. we optimize each
infinitesimal contribution to $\disp \tilde f$ that comes from a small vicinity around some
$(a_0, \bmath b_0)$. In this case, the integral~(\ref{Rfunc}) is approximated by just the
naked integrand:
\begin{align}
R_{\mathcal S}(\bmath x,\bmath x') \propto \frac{1}{C_{\psi\gamma}} \psi\left(\frac{\bmath x - \bmath b}{a}\right) \gamma\left(\frac{\bmath x' - \bmath b}{a}\right) \implies \nonumber\\
\implies \| R_{\mathcal S} \| \propto \frac{\|\gamma\|}{|C_{\psi\gamma}|},
\end{align}
where the omitted all coefficients that do not depend on $\gamma$.

$C_{\psi\gamma}$ is basically a scalar product of two functions, $\hat\psi(s)/s^n$ and
$\hat\gamma(s)$:
\begin{equation}
C_{\psi\gamma} = \int\limits_0^{+\infty} \frac{\hat\psi^*(s)}{s^n} \hat\gamma(s) s^{n-1} ds = \frac{1}{S_n} \int\limits_{\mathbb R^n} \frac{\hat\psi^*(\bmath s)}{|\bmath s|^n} \hat\gamma(\bmath s) d\bmath s,
\end{equation}
By the Plancherel theorem, this scalar product can be equivalently determined by an
integral in the $\bmath t$-space, operating on the corresponding Fourier originals.
Therefore, to minimize $\| R_{\mathcal S} \|^2$ with respect to $\gamma$, we must maximize
the ratio $(\hat\psi/s^n, \hat\gamma)^2/\|\hat\gamma\|^2$. Since this ratio is a
cosine-angle type quantity, the obvious maximum is achieved for
\begin{equation}
\hat\gamma(s) \propto s^{-n} \hat\psi(s) = - s^{2-n} \hat\varphi(s), \quad s\geq 0.
\label{gammaopt}
\end{equation}

Here we used the property that applying the Laplace operator to some $\varphi(\bmath t)$ is
equivalent to multiplying $\hat\varphi(\bmath s)$ by $-|\bmath s|^2$ in the Fourier
$s$-space.

From~(\ref{gammaopt}) one can derive a remarkable result: in the 2D case ($n=2$) the
optimal reconstruction kernel coincides with the negated generating function, $\gamma(t)
\equiv -\varphi(t)$.

If $\gamma$ is computed from~(\ref{gammaopt}) without any additional coefficients then the
inversion constant in~(\ref{icwtiso}) becomes
\begin{equation}
C_{\psi\gamma} = \int\limits_0^{+\infty} |\hat\varphi(s)|^2 s^{3-n} ds = \frac{1}{S_n}\int\limits_{\mathbb R^n} |\hat\gamma(\bmath s)|^2 d\bmath s.
\label{Copt}
\end{equation}
This is basically the $L_2$ norm of $\gamma$, or its total Fourier power.

When computing such an integral for a linear mixture $\varphi = \sum_\nu c_\nu
\varphi_\nu$, where $\varphi_\nu$ are defined via~(\ref{psc}), one may utilize the
following useful identity:
\begin{equation}
\int\limits_0^{+\infty} \hat\varphi_\nu(s) \hat\varphi_\mu(s) s ds = \frac{4\pi^2}{1+\nu+\mu},
\end{equation}
which assumes $n=2$.

The general model~(\ref{wmp}) is not used here, but for the 2DMHAT case it is easy to
obtain $C_{\psi\gamma} = 2\pi^2$.

In addition it is necessary to take into account the normalization scaling~(\ref{evennrm}),
which implies $\gamma_{\rm nrm}(t) = K k^{2-n} \gamma(kt)$ and requires $C_{\psi\gamma}$ to
be scaled up by $K^2 k^{4-3n}$.

In particular, for the classic 2DMHAT and for our new 2DOPT2 wavelet, the result is
\begin{align}
\text{2DMHAT} &: \quad C_{\psi\gamma} = 0.922636, \nonumber\\
\text{2DOPT2} &: \quad C_{\psi\gamma} = 0.875171.
\end{align}

Summarizing this section, we note that this method of finding an optimal reconstruction
kernel seemingly works only for $n=1$, $n=2$, and possibly for $n=3$. Already for $n=3$ we
obtain a singular Fourier image $\hat\gamma = \hat\varphi/s$, which looks rather
impractical, though probably still tractable mathematically. But for $n=4$ and above, the
constant $C_{\psi\gamma}$ becomes infinite in~(\ref{Copt}). Therefore, the cases of higher
dimensions still need a more careful investigation.

\section{Reconstruction of the p.d.f. after wavelet denoising}
\label{sec_recon}
We adopt the same method of the p.d.f. reconstruction as we used in the 1D case. This is
basically the matching pursuit algorithm, intended to model the p.d.f. $f(x)$ in the most
economic way, still agreeing with the significance threshold, that is
avoiding~(\ref{zthrmax}) in the normality domain $\mathcal D$. This is based on an
iterative application of noise filtering based on~(\ref{zthrmax}) and application of the
inversion formula~(\ref{icwtiso}), until no signficant structures are left in $\mathcal D$.

This means that all the steps~(\ref{ftf}) are iterated until the following
condition is satisfied:
\begin{equation}
\max_{\mathcal D} |z(a,\bmath b)| \leq z_{\rm thr}.
\label{Linf}
\end{equation}
Hence, we basically solve a type of a nonlinear programming task considering the $L_\infty$
norm of $z$ as a constraint, while some implicit simplicity measure of $Y$ as an objective
function. The latter objective can be formulated as to minimize the number of nonzero
wavelet coefficients that contribute to $\tilde f$ via~(\ref{icwtiso}), so in some sense
this might be treated as an $L_0$ norm in the $(a,\bmath b)$-space \citep{Hara17}. That is,
we try to obtain the simplest model of $f$ still satisfying the noise significance limits.

In fact, any other combination of $L_p$ norms is possible here, e.g. \citet{McEwen17} used
a $\chi^2$ metric in place of~(\ref{Linf}), so this implies the $L_2$ norm. Each
combination of norms may have its own advantages or disadvantages, but this discussion is
out of the scope of this paper.

More details of this inversion method are already given in \citep{Baluev18a} and disscussed
in \citep{Baluev18b}, and here we only generalize it to the 2D case in an obvious manner.

The associated computation complexity is, of course, significantly larger in the 2D case.
Unfortunately, current implementation of this part is relatively impractical: processing of
the asteroid test case presented below was possible only with a top-class multi-cpu system.
One possible solution to further soften the `dimensionality curse' and achive a practically
feasible code might be to integrate~(\ref{icwtiso}) by using the Monte Carlo approach or by
applying genetic algorithms, instead of relying on an inefficient regular grid. However, we
did not assess or implement this option yet.

In any case, we emphasize that the reconstruction of the p.d.f. is not the main goal of our
analysis. It carries more demonstrative purpose, while the most important part is the
detection of significant patterns in the CWT, i.e. the detection of structures that passes
the threshold~(\ref{zthrmax}).

\section{The updated code: improved 1D and new 2D algorithm}
\label{sec_code}
The C++ code with a command-line analysis pipeline is available for download at
\texttt{https://sourceforge.net/projects/waveletstat}.

This package now contains algorithms for the both 1D and 2D analysis. The 1D code published
with the previous paper, is now considerably revised and improved. The changes were worked
out after a year of practical application and tests of the code. They include the
following.

\begin{enumerate}
\item The computing speed was improved a lot, thanks to optimizations of the algorithm
itself, as well to an efficient parallelization, based on the {\sc C++11} multithreading
tools (the previous version was only partly parallelized). We also optimized the code in
such a way that it is now capable to handle large samples that involve $\sim 10^5$ of
objects and above (like e.g. asteroid databases).

\item Output files were optimized for size. Previously, the file with a wavelet transform
could grow extremely huge (e.g. $\sim 10$~Gb and beyond). Now we revised its format by
removing unnecessary debug-like information, portions outside of the Gaussian domain, and
values below some `compression' threshold (it can be quite a mild, e.g. $z=1$ is alllows
for a quite efficient compression). The resulting file size was reduced by the factor of
$\sim 100-1000$, depending on the parameters.

\item Now user can select hard as well as soft thresholding when reconstructing the p.d.f.
(previously only the soft thresholding was available). The hard thresholding appears
considerably faster thanks to a smaller number of matching pursuit iterations, though soft
thresholding results in a more smooth p.d.f. model.

\item The algorithm may now adopt three types of the initial p.d.f.: constant zero (an
improper distribution with infinite variance), flat in a specified analysis range, and
Gaussian which mean and variance are estimated from the sample. Previously only the zero
initial p.d.f. was allowed. The Gaussian model allows to adaptively incorporate most of the
large-scale p.d.f. structure, avoiding reconstructing it by iterations. Hence, the p.d.f.
reconstruction process runs faster, since only the small-scale details are iterated. The
formula used to compute the CWT of a Gaussian or of a flat distribution are given in
\ref{sec_gf}. The flat model is natural for angular variables (in the 1D case). This is an
alternative (and possibly more justified) approach then the workaround used by
\citet{Baluev18b}, which was based on sample cloning to get rid of the p.d.f. jumps near
the segment endpoints.

\item The normality domain $\mathcal D$ is now considered differently at small scales
($a<\disp x$) and at large scales ($a>\disp x$). In the small-scale zone we require the
following conditions to be satisfied for a given point $(a,\bmath b)$: (i) number of sample
points $x_i$ within the wavelet cut-off domain (see below) is at least $4$, (ii) the
quantity $n(a,\bmath b)$ is at least $1$, (iii) the variance estimate $\widetilde
D(a,\bmath b)$ is non-zero, and finally (iv) the primary normality
criterion~(\ref{nrmtest}), or its 1D-case version from \citep{Baluev18a}, is satisfied. The
first three additional conditions were necessary to workaround several technical issues
revealed in the practical application of the algorithm with practical data. In the
large-scale zone, we require that (i) $\widetilde D(a,\bmath b)$ is non-zero, and (ii)
$n(a,b)$ is above $n_{\rm thr} = \varepsilon^{-2}$, typically $10$ (small $n$ indicates a
small number statistic). Such a special treatment of the large-scale zone appeared because
in practice the normality test~(\ref{nrmtest}) is often failed at large scales, but
nonetheless the values of the $z$ statistic are always large there, so this non-Gaussianity
is usually unrelated to the effect of small-sample statistic and does not indicate that the
corresponding CWT values are insignificant. The large-scale zone contains information about
only a global structure of $f(\bmath x)$, for example about its unit normalization.
Although this global-scale structure usually is not very interesting in itself, it is still
important to accurately reconstruct it. Therefore, we decided not to use~(\ref{nrmtest}) at
large scales.

\item The normality domain determined in accordance with the instructions above is further
postprocessed to remove the extremely porous structure of its boundary at small scales. In
the 1D case, we purge away all the points in the $(a,b)$-plane that were formally
classified as Gaussian, but appear isolated (surrounded by non-Gaussian points at the
neighbouring $b$), or if the continuous segment formed by Gaussian points is too short
along the $b$-axis (shorter than $0.3 a$). All such isolated points or too short point
sequences are reclassified as non-Gaussian. This allowed to make the Gaussian/non-Gaussian
transition boundary more regular, hence to reduce the size of the output and of the
associated figure files. In the 2D case, we apply a more intricate algorithm based on a
disk-kernel dilation of the non-Gaussian domain in the $\bmath b$-plane, followed by a
similar dilation of the remaining Gaussian domain. This resulted in the removal of all
Gaussian pieces that could not embed a disk of a given size\footnote{This method was
inspired by the `morphology' operator of the {\sc ImageMagick} utility. See the associated
web page for further explanation, \texttt{http://www.imagemagick.org/Usage/morphology/}.
Our approach is equivalent to applying the `\texttt{Open}' method with a `\texttt{Disk}'
kernel.}. The radius of the disk is set so that its area appears equal to $0.3 a^2$.

\item It appeared that the previous version of the code missed many of the terms that
should appear in the normality test (hidden inside $q_{1,-1}$ and $q_{2,-6}$), due to a
programming bug. Now we fixed it, although this fortunately did not have practical
consequences on the results, because the missed terms appeared negligible in our tests.
Notice that in the 2D version of the algorithm all the terms $q_{2,j}$ for any $j$, as well
as all $q_{1,j}$ with $j>-3$, are neglected in any case. This error significantly affected
only the coefficient $W_{02}$, ecpecially for the WAVE2 wavelet in the 1D case. However,
this term contributes to the FAP with a small factor $z^{-2}$, so it usually remains small.
In the 2D case, the coefficient $W_{02}$ is neglected at all.

\item Another potentially important mistake in the legacy code appeared when estimating
various sample momenta by the corresponding sample averages. For example, the sample-based
estimator for the covariance $\lambda_{XY} = \cov(X,Y)$ is
\begin{align}
l_{XY} = \frac{1}{N} \sum_{i=1}^N (X_i-l_X) (Y_i-l_Y), \nonumber\\
l_X = \frac{1}{N} \sum_{i=1}^N X_i, \quad l_Y = \frac{1}{N} \sum_{i=1}^N Y_i.
\label{covsum}
\end{align}
In practice the quantities $X$ and $Y$ are related to the wavelet $\psi$ or to its
derivatives, computed in the point $(x-b)/a$. Since all such functions are well-localized,
with exponential tails in the 1D case, we optimized the code by neglecting
in~(\ref{covsum}) all the quantities for which $|(x-b)/a|$ is above some limit. Denoting
such reduced sum by ${\sum}'$, and the number of the non-negligible terms by $N'$, this
would result in:
\begin{align}
l_{XY} &\approx \frac{1}{N} {\sum}' (X_i-l_X) (Y_i-l_Y) + \frac{N-N'}{N} l_X l_Y, \nonumber\\
& l_X \approx \frac{1}{N} {\sum}' X_i, \quad l_Y \approx \frac{1}{N} {\sum}' Y_i.
\label{csred}
\end{align}
However, in the old version of the code the second term of~(\ref{csred}), containing $l_X
l_Y$, was mistakenly missed. Fortunately, this term is only important at large scales,
because otherwise the reduced sum ${\sum}'$ contains relatively few terms. Then $l_X$ and
$l_Y$ become small quantities of the order $N'/N$, and so is $l_{XY}$. So the missed
addition $l_X l_Y$ has the relative magnitude of $\sim N^{-1}$. A similar mistake appeared
in the momenta of the third degree:
\begin{align}
l_{XYZ} &= \frac{1}{N} \sum_{i=1}^N (X_i-l_X) (Y_i-l_Y) (Z_i-l_Z) \nonumber\\
&\approx \frac{1}{N} {\sum}' (X_i-l_X) (Y_i-l_Y) (Z_i-l_Z) \nonumber\\
&- \frac{N-N'}{N} l_X l_Y l_Z,
\end{align}
where the last term was again missed in the old code. However, here it has the relative
magnitude of just $N^{-2}$. Analogously, for the quad-degree momenta $l_{XYZT}$ the
corresponding addition would be about $N^{-3}$ relatively to the total sum. After fixing
this bug, we did not notice any visible changes in practical tests.

\item The published algorithm now involves a 2D version, which uses all the theory
described above. However, the 2D version may be still be considered as beta, since it got
smaller amount of testing. The code is organized universally in the sense that the 1D and
2D algorithms are produced by the same source files, just compiled with different
parameters. In the future such code organization may help us to extend the algorithms to
even larger dimensions.
\end{enumerate}

Now let us give some more details concerning the computation algorithm that we omitted in
\citep{Baluev18a}. First of all, it takes the following input.
\begin{enumerate}
\item The input sample $\{x_i\}$.
\item The desired wavelet ($4$ wavelets available in the 1D case, and $2$ in the 2D case).
\item The desired shift range $[b_{\min}, b_{\max}]$, or the corresponding square domain in
the 2D case (also determined by just $b_{\min}$ and $b_{\max}$).
\item The scale range $[a_{\min}, a_{\max}]$, where $a_{\min}$ must be a small positive
value, but $a_{\max}$ can be infinite, which would mean $\kappa_{\min}=0$. The reasonable
value of $a_{\min}$ depends, in general, on the sample size $N$ (larger $N$ means smaller
$a_{\min}$) and on the peaky parts of $f(x)$ (where $x_i$ are sampled more densely, hence
the most small scales can be reached). However, we cannot give any mathematically definite
formula for $a_{\min}$, so this parameter needs to be guessed by the user or determined by
experimenting. Reducing $a_{\min}$ would slow the computation down, but increasing
$a_{\min}$ too much would result in the lost CWT structures. Usually there is some `safe'
$a_{\min}$ that could allow to discover all significant structures in the $(a,b)$-space
(below this $a_{\min}$ all values of $z(a,b)$ are smaller than the noise threshold).
\item The adimensional parameter $h$ that defines the resolution of various grids.
Basically, it has the meaning of how fine we intend to sample $\psi(t)$.
\item Other fine-tuning options, in particular the type of the initial p.d.f. approximation
$Y_0$.
\end{enumerate}

In the output, our algorithm first computes and saves the initial SWT and auxiliary data,
namely $Y(a,b)-Y_0(a,b)$ and $z(a,b)$, the simplified normality indicator $n(a,b)$, and the
logical (0/1) flag that indicates whether the full normality test was passed or not. Some
additional more detailed output can be requested if necessary.

On the second phase, the algorithm reconstructs the p.d.f. model by matching pursuit
iterations, and filtering out all points that fail the significance test for $|z(a,b)|$.
This is done simultaneously for three probabilistic levels: $\FAP=31\%$ (or $1$-sigma
level), $\FAP=4.6\%$ (or $2$-sigma level), and $\FAP=0.27\%$ (or $3$-sigma level). In the
end, the reconstructed p.d.f. models corresponding to these three tolerance levels is
saved, as well as the corresponding residual SWTs (so that the user can verify that no
significant structures were left in the $(a,b)$-plane after the iterations). This second
phase takes usually considerably longer time that just to compute the SWT. The matching
pursuit iteration are continued until the volume of the domain where~(\ref{Linf}) is
violated, is reduced to $0.3$ in the small-scale zone and $0.5$ in the large-scale zone, or
until $100$ iterations are made.

The algorithm involves several optimizations, in particular it keeps only such terms in the
summation~(\ref{wavest}), and similar summations, for which the argument of $\psi(t)$
appeared below some maximum radius. This basically cuts away the tails of $\psi(t)$,
removing negligible values from the summations. A similar cutting is applied when
integrating~(\ref{icwtiso}) with the reconstruction kernel $\gamma(t)$. In the case of
2DOPT2 wavelet there is a natural limit on $t$, where $\psi(t)$ vanishes by definition. In
the 1D case, the limit is selected to be $5-7$, depending on the wavelet.

The sampling of the SWT with respect to the parameter $b$ can be linear or logarithmic,
depending on the user request. In the latter case, we basically analyse the variable $\log
b$ instead of $b$, and regarding the initial linear $b$, the scale $a$ attains the meaning
of the relative scale.

With respect to the parameter $a$, the SWT is always sampled using a special non-linear
scaling. It is such that the following quantity appears sampled linearly (i.e.,
equidistantly):
\begin{equation}
l(a) = \log\left(1 + \frac{\sigma}{a}\right) = \log(1 + \sigma \kappa),
\label{linlog}
\end{equation}
where $\sigma$ is the input sample variance. This formula was motivated by our intention to
sample local peaks of the SWT approximately uniformly everywhere in the domain $\mathcal
D$. From the formula~(\ref{W00Tnrm}) and from the behavior of individual components in the
gradient $z'$, it follows that the local peaks of $z(a,\bmath b)$ should have approximately
constant width in terms of the variables $\log a$ (for small $a$) and $\bmath c$. The
logarithmic natural scaling for $a$ also follows if we consider the integral
autocorrelation function for $\psi((b-x)/a)$ with different $a$ and $b$: it depends on the
difference of $c=-b/a$ and on the \emph{ratio} of the scales $a$. But for large $a$ we face
a degeneracy, so a more reasonable variable becomes $\kappa = 1/a$. Therefore, the
nonlinear parametrization~(\ref{linlog}) appeared as the mixture of the two scales.

Our algorithm contains two major phases: the computation of the SWT $Y(a,b)$, and the
iterative reconstruction of the denoised p.d.f. $f(x)$, using the inversion
formula~(\ref{icwtiso}). The second phase can be switched off by setting an appropriate
command argument. In this case, the algorithm will not only process faster, but also use a
considerably smaller amount of memory, since it does not need to store the entire CWT
array.

The computational complexity of the first phase, as implemented in our algorithm, can be
roughly expressed by a formula
\begin{equation}
N_\psi \sim h^{-n-1} N L^n \Delta l,
\label{nc1}
\end{equation}
where $N_\psi$ is the number of wavelet evaluations, $n$ is the dimension ($1$ or $2$), $h$
is an adimensional grid resolution parameter (about $0.03$ to $0.1$), $N$ is the input
sample size, $L$ is an adimensional localization range of the wavelet (depending on the
wavelet, varies from $\sim 4$ to $\sim 10$), and $l$ depends on the requested scale range
$[a_{\min},a_{\max}]$:
\begin{equation}
\Delta l = l(a_{\min})-l(a_{\max}).
\end{equation}
If $a_{\rm max}=\infty$ (which is a valid value actually meaning $\kappa_{\min}=0$), this
$\Delta l$ turns into just $\log(1+\sigma/a_{\min})$. Remarkably, $N_\psi$ does not depend on $f(x)$ or
on the requested range of the $b$ parameter.

The second phase is iterative, and each iterations contain (i) the inversion of the
thresholded SWT and (ii) an update of the CWT comparison model $Y_0$. The complexity of the
first part is roughly expressed as
\begin{equation}
N_\gamma \sim h^{-2n-1} \left(\frac{BL}{a_{\min}}\right)^n \Delta l S_z,
\label{nc21}
\end{equation}
where $N_\gamma$ is the number of evaluations of $\gamma(t)$, $B$ is the range of the $b$
parameter requested by user (in the 1D case), or the side of the corresponding square (in
the 2D case). The complexity also depends on the fraction $S_z$ of the statistically
significant values of $z$-statistic that remained at the current iteration. This is because
insignificant values do not contribute in the result ($S_z\ll 1$ usually). Remarkably,
$N_\gamma$ does not depend on $f(x)$ or on the input sample size $N$ directly, though small
$N$ would imply that reasonable $a_{\min}$ is likely large.

The second part of the second phase is equivalent to computing an integral over $n$
dimensions, for parameters covering a grid in $n+1$ dimensions. Taking into account all the
optimizations, this corresponds to the following numeric complexity:
\begin{equation}
N_\psi \sim h^{-2n-1} \left(\frac{BL}{a_{\min}}\right)^n \Delta l.
\label{nc22}
\end{equation}
Compared to~(\ref{nc21}), this part of the algorithm does not profit from the small factor
$S_z$. Basically, it just plainly applies the CWT to the current model of $f(x)$. Further
on, this can be optimized by using the information that we should update the CWT in only a
small domain surrounding the `significant' part of the $(a,b)$-grid, because changes in the
`insignificant' part should be mostly negligible (thanks to the wavelet orthogonality).
This would allow to add the factor $S_z$ to~(\ref{nc22}). However this is still a work to
do.

As we can see, the p.d.f. reconstruction phase of the algorithm currently has rather poor
efficiency, especially for $n=2$, so it needs to be optimized further to become more
practical. But concerning the case $n=1$, both phases of our algorithm appear practically
feasible, even for very large samples containing $N\sim 10^6$ objects.

Further technical details can be found in the \texttt{readme} file supplied with the code.

\section{Practical tests and examples}
\label{sec_tests}
\subsection{Monte Carlo tests of the significance levels}
Approximating the false alarm probability was one of the most complicated and possibly
vulnerable part of the method. Therefore, we performed Monte Carlo simulations to verify
that the relevant formulae is adequate and works as expected.

\begin{figure*}[!t]
\includegraphics[width=0.99\textwidth]{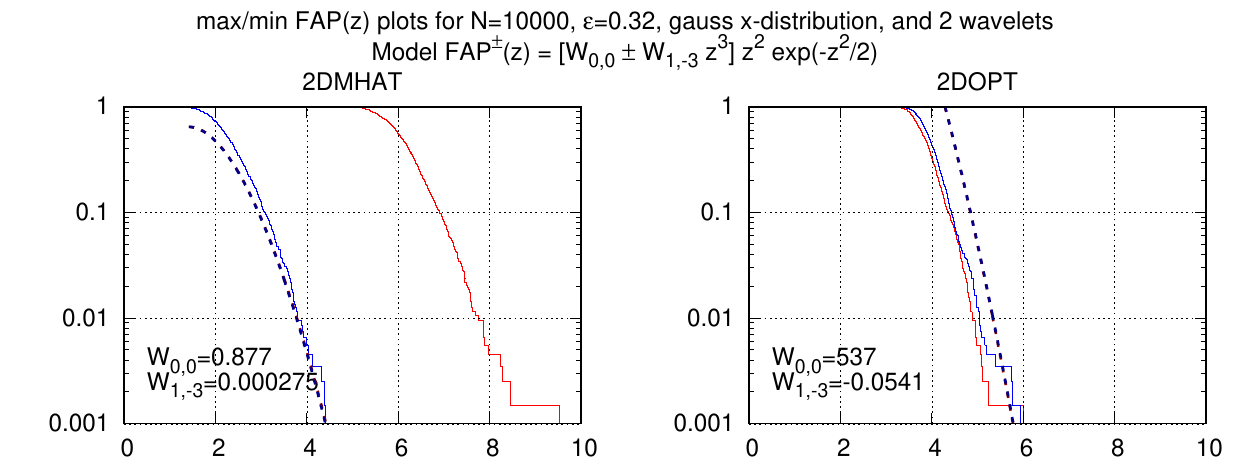}
\caption{Analytic approximation~(\ref{FAPgauss}), compared with Monte Carlo simulations.
The simulations involved the normality criterion~(\ref{nrmtest}) and thus the coefficients
$W_{00}$ was restricted to the normality domain $\mathcal D$. Red-colored curves refer to
positive extrema distributions (maxima of $z(a,b)$, $\FAP^+$), while blue-colored ones are
for the negative ones (minima of $z(a,b)$, $\FAP^-$), though they appear mostly
indistinguishable from each other. Theoretic approximations are shown in thicker dashed
lines.}
\label{fig_evdgauss}
\end{figure*}

However, our simulations could only be limited, because the computing the CWT on a 3D grid
is a practically hard task. Moreover, we had to increse the size of the simulated sample
from $N\sim 1000$ (in 1D) to $N\sim 10000$ (in 2D), because otherwise the CWT would contain
mainly the large-scale structure, which is not very interesting to test.

Therefore, the number of Monte Carlo trial was reduced to only $1000$. In each trial we
simulated a random sample, drawing it from the standard 2D Gaussian distribution, $\mathcal
N_2(\mu=0,\sigma^2=1)$. Then we computed the CWT, cutting all the scales below
$a_{\min}=\sigma/\sqrt{N}$, and simultaneously determined the normality domain $\mathcal
D$. Then we computed a pair of extreme values in $\mathcal D$:
\begin{equation}
z_{\max}^+ = \max_{\mathcal D} z(a,\bmath b), \quad
z_{\max}^- = \max_{\mathcal D} [-z(a,\bmath b)].
\end{equation}
To accurately compute these quantities, for each candidate extrema we applied a Newtonian
iteration scheme in the 3D space $(a,\bmath b)$. The necessary 3D gradient and $3\times 3$
Hessian matrix were computed simultaneously with $z$, using analytic formula (we omit
them). This approach allowed us to reduce the computation time by rarifying the 3D grid,
but without reducing the resulting values of $z_{\max}^\pm$.

The simulated extrema $z_{\rm max}^\pm$ then were used to construct their empirical
distribution functions and hence the simulated $\FAP$ curves. The latters were compared
with the analytic approximations~(\ref{FAPgauss},\ref{FAPngauss}). This comparison is shown
in Fig.~\ref{fig_evdgauss}. We can see that for the 2DOPT2 wavelet the agreement is good,
i.e. our analytic formulae allow to determine $z$-threshold with a quite stisfactory
accuracy. Moreover, we can see that the simulated distributions of $z_{\max}^\pm$ almost
coincide with each other, so this wavelet does not generate any significant skewness.

However, the 2D MHAT wavelet generated very small normality domain (indicated by small
$W_{00}$). And even in this reduced domain the behaviour of $z$ seems not quite Gaussian,
because the distributions of $z_{\max}^\pm$ appear pretty different from each other. This
could appear because in the 2D case our normality test neglects several higher-order terms
that likely work here (even if the skewness $\As y$ is formally suppressed). The analytic
$\FAP$ approximation becomes expectedly poor in this case.

\subsection{A practical test case: the 2D eccentricity distribution in the Main Belt}
The 2D wavelet analysis is obviously more demanding, not only with respect to the required
computing capabilities, but also with respect to the task and input data.

Let us consider a random sample of size $N$, distributed more or less uniformly in the
$n$-dim cube having the side $L$. Then the average density of the points would be $N/L^n$,
while the typical size of a volume containing just a single point would be $a_{\min} \sim L
N^{-1/n}$. Therefore, volumes sized as $\sim a_{\min}$ or below would not be sampled
adequately, so $a_{\min}$ represents a natural smallest scale that can be analysed. In the
1D case we had $a_{\min} \sim L/N$, so it could be quite small for large samples. But now
$a_{\min} \sim L/\sqrt N$, so even rather large samples do not allow to perform a
fine-resolution analysis of the underlying p.d.f.

For example, having $N\sim 1000$ objects in our sample we cannot go below $a_{\min} \sim
L/30$, where $L$ is about sample variance. This is relatively moderate scale limit,
compared to the 1D case, when we would have $a_{\min}\sim L/1000$ with the same $N$.
Therefore, to reveal the full power of our method, we have to work with very large samples.

We opted to deal with the asteroid catalog \texttt{astorb.dat} of the Lowell
observatory.\footnote{See url ftp://ftp.lowell.edu/pub/elgb/astorb.html.} For our goals we
selected only numbered asteroids, which gave us $N\sim 5\times 10^5$. In theory, this would
allow us to reach the scales as small as $a_{\min} \sim L/700$, which is comparable to
$N=700$ in the 1D case.

Since our analysis is isotropic, we need to select two physically and numerically
comparable parameters. Initially we decided to use for this goal eccentric parameters
$(e\cos\varpi,e\sin\varpi)$, where $e$ is orbital eccentricity of an asteroid, while
$\varpi$ is longitude of the perihelion. However, after some thinkering we decided that the
parameter $\varpi$ is not entirely physical, because it is defined as a sum of two angles
lying in different planes. We therefore stopped on the following pair of quantities:
\begin{align}
f_x' &= \frac{e f_x}{\sqrt{f_x^2+f_y^2}} &= e \cos\varpi', \nonumber\\
f_y' &= \frac{e f_y}{\sqrt{f_x^2+f_y^2}} &= e \sin\varpi',
\end{align}
where $f_x$ and $f_y$ are Cartesian components of the Laplace vector. This is a vector of
length $e$ directed from the Sun to orbital perihelion, and based on the formulae of
two-body motion \citep{Kholsh-twobody} it can be expressed as follows:
\begin{align}
\bmath f = e \{&\cos\omega\cos\Omega-\cos i\sin\omega\sin\Omega, \nonumber\\
&\cos\omega\sin\Omega+\cos i\sin\omega\cos\Omega, \nonumber\\
&\sin i\sin\omega\, \},
\end{align}
where $\omega$ is argument of the perihelion, $\Omega$ is the ascending node longitude, and
$i$ is orbital inclination.

Clearly, if $i$ is small then $f_x' \simeq e\cos\varpi$ and $f_y' \simeq e\sin\varpi$, so
we obtain almost what we initially aimed to obtain. However, for large $i$ these $f_x'$ and
$f_y'$ might be slightly different, representing in general the normalized projection of
the Laplace vector on the ecliptic plane.

\begin{figure*}[!t]
\begin{tabular}{cc}
\includegraphics[width=0.49\linewidth]{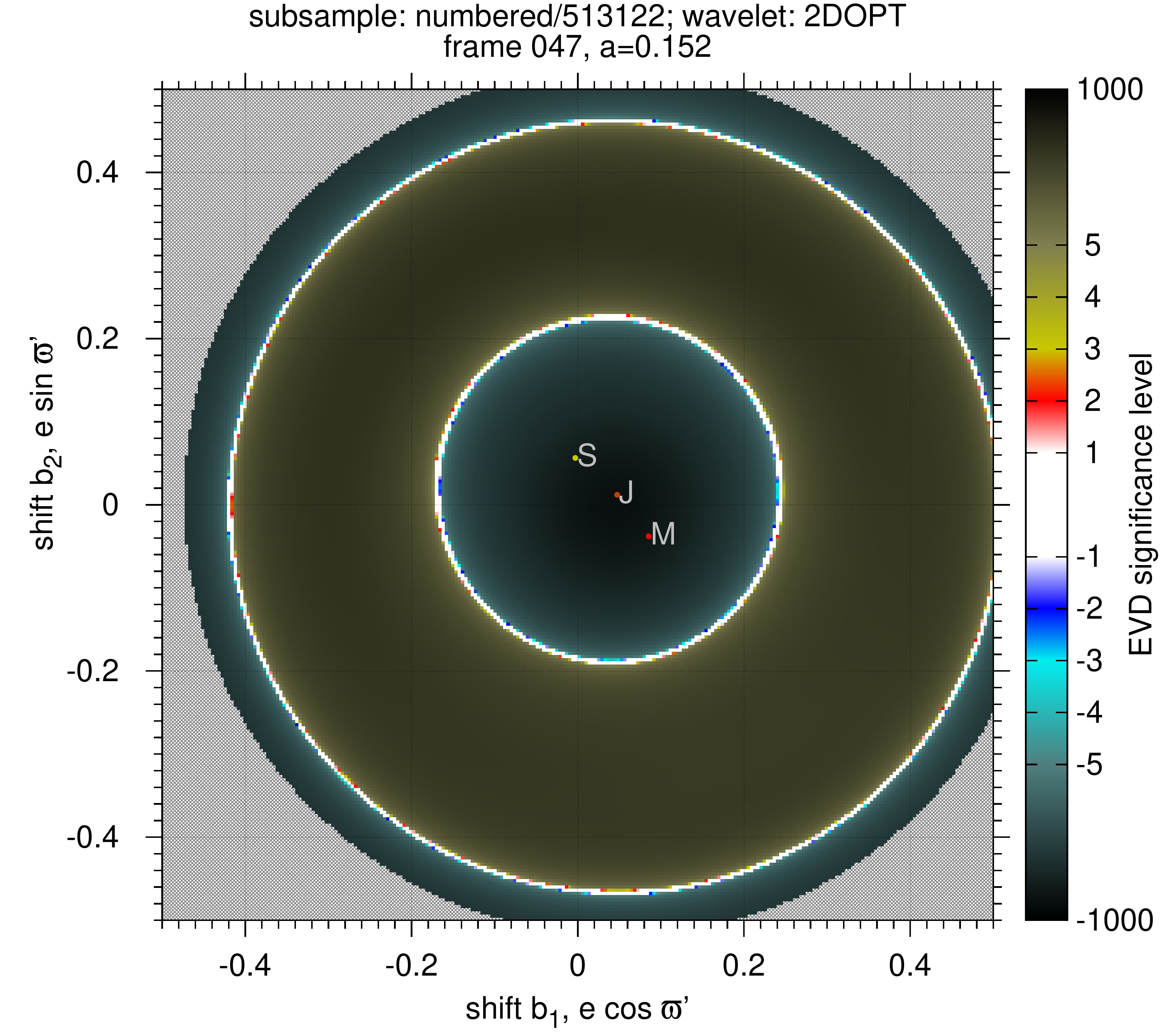} &
\includegraphics[width=0.49\linewidth]{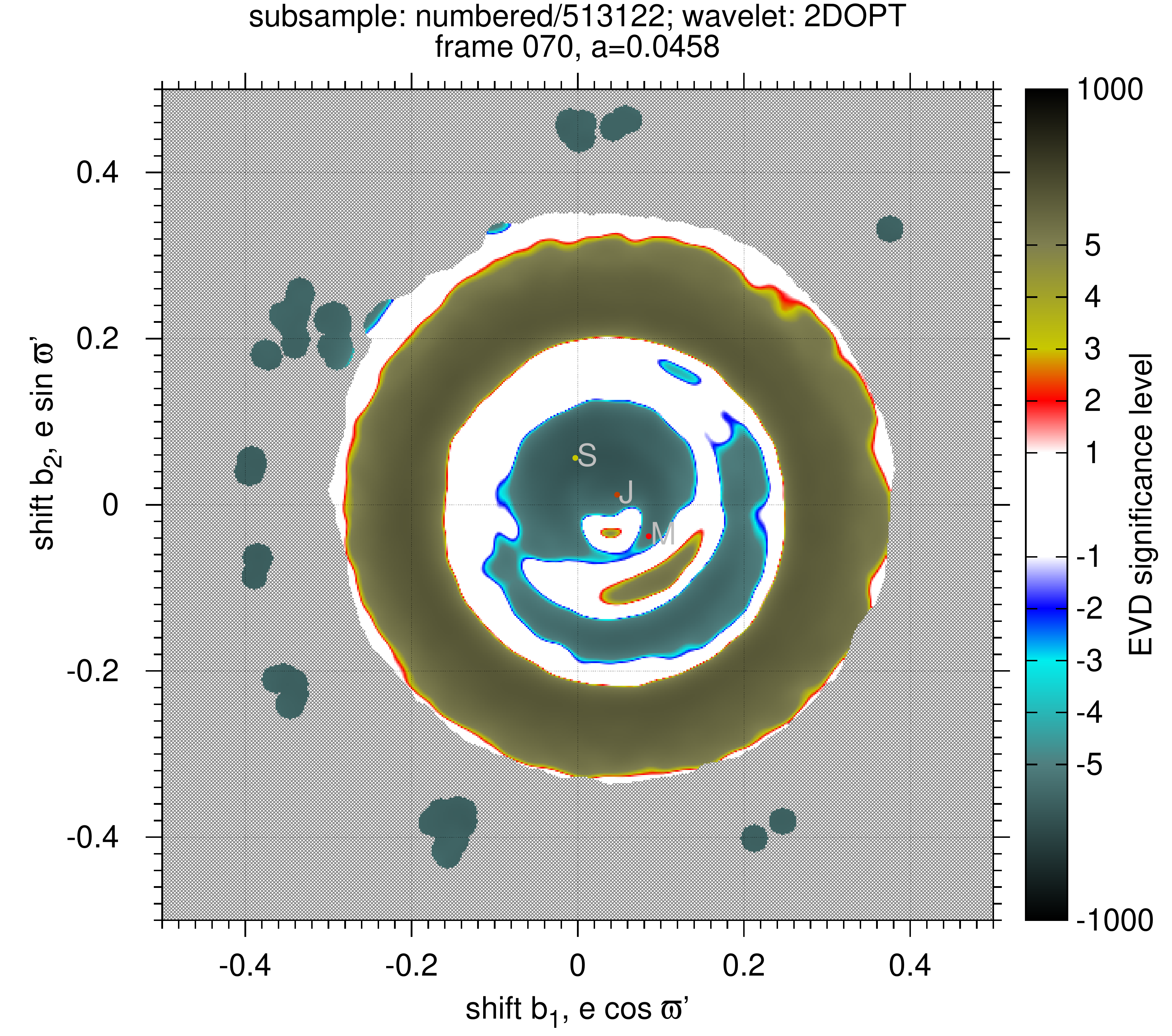} \\
\includegraphics[width=0.49\linewidth]{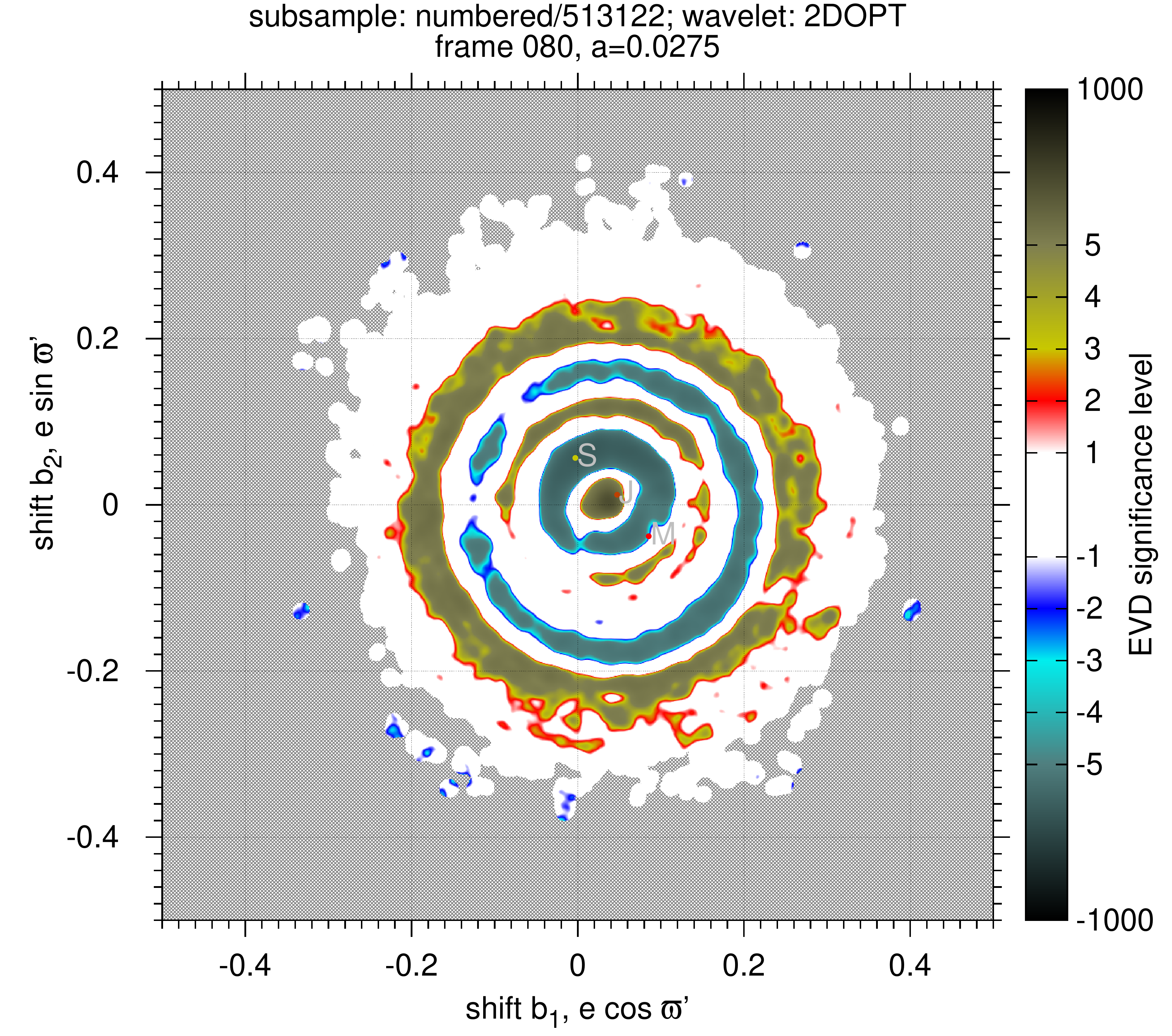} &
\includegraphics[width=0.49\linewidth]{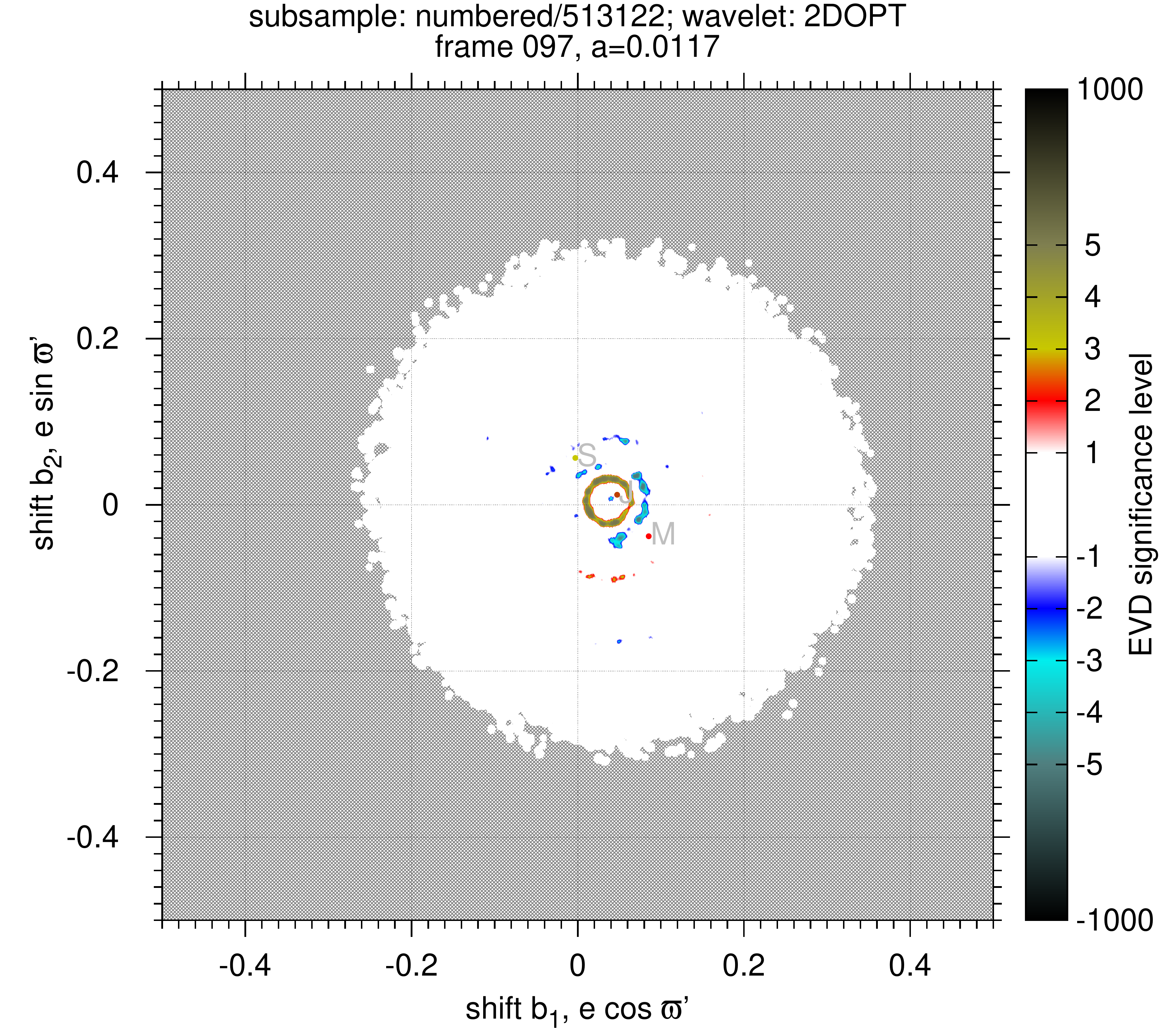}
\end{tabular}
\caption{Several slices of the CWT for the 2D distribution of $(e\cos\varpi',
e\sin\varpi')$ for the Main Belt asteroids. Slices correspond to several selected scales
$a$, decresing from $0.152$ (approximately the sample variance) to $\sim 0.01$. The color
corresponds to the significance of the corresponding $z(a,\bmath b)$, as transformed to
normal quantiles, with sign symbolically reflecting the sign of $z$. The hashed regions of
each plot correspond to the non-Gaussian domain, where~(\ref{nrmtest}) was failed. In each
frame we also label the position of Mars, Jupiter, and Saturn (the three main perturbers).}
\label{fig_astecc}
\end{figure*}

\begin{figure*}[!t]
\begin{tabular}{cc}
\includegraphics[width=0.49\linewidth]{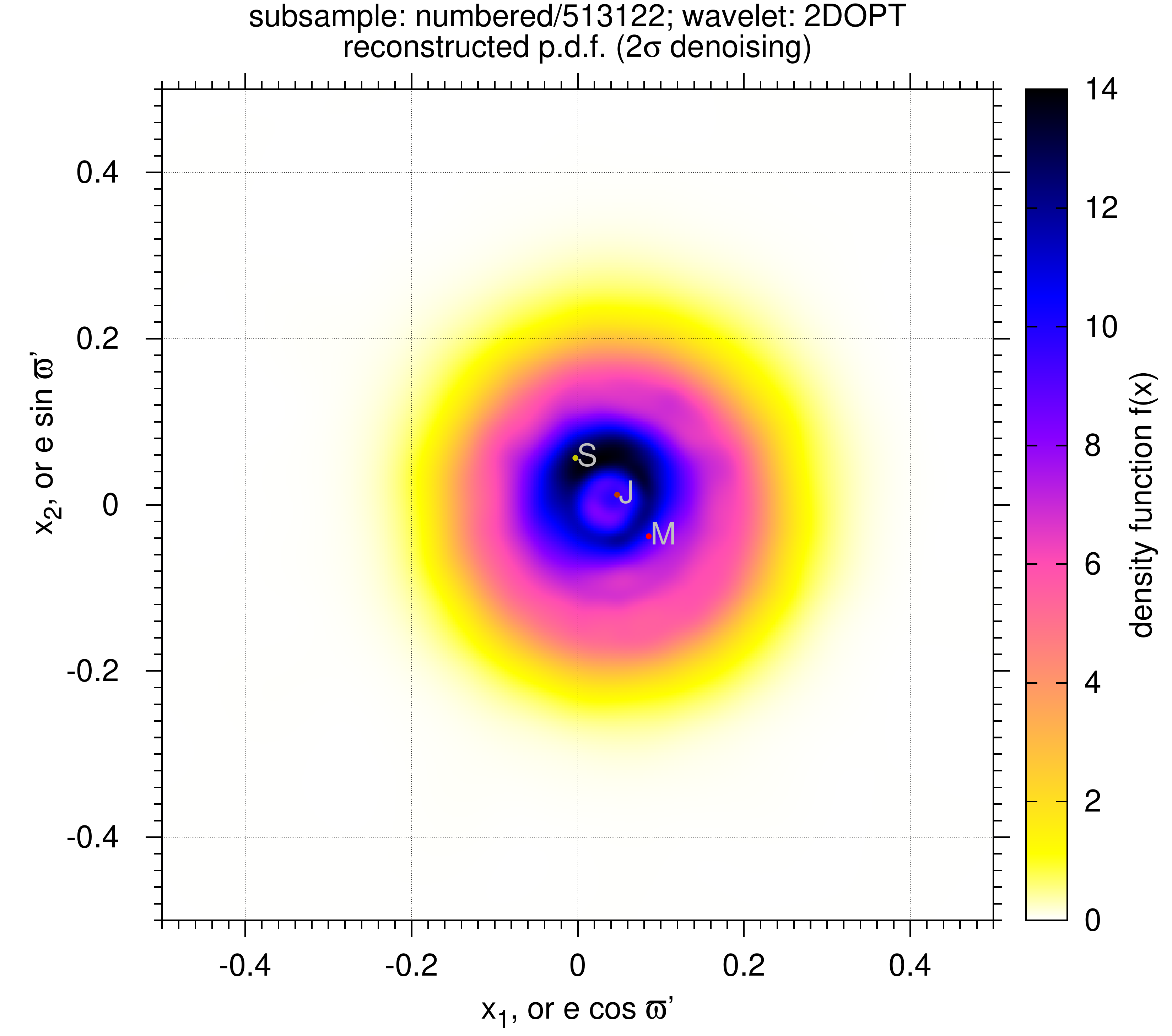} &
\includegraphics[width=0.49\linewidth]{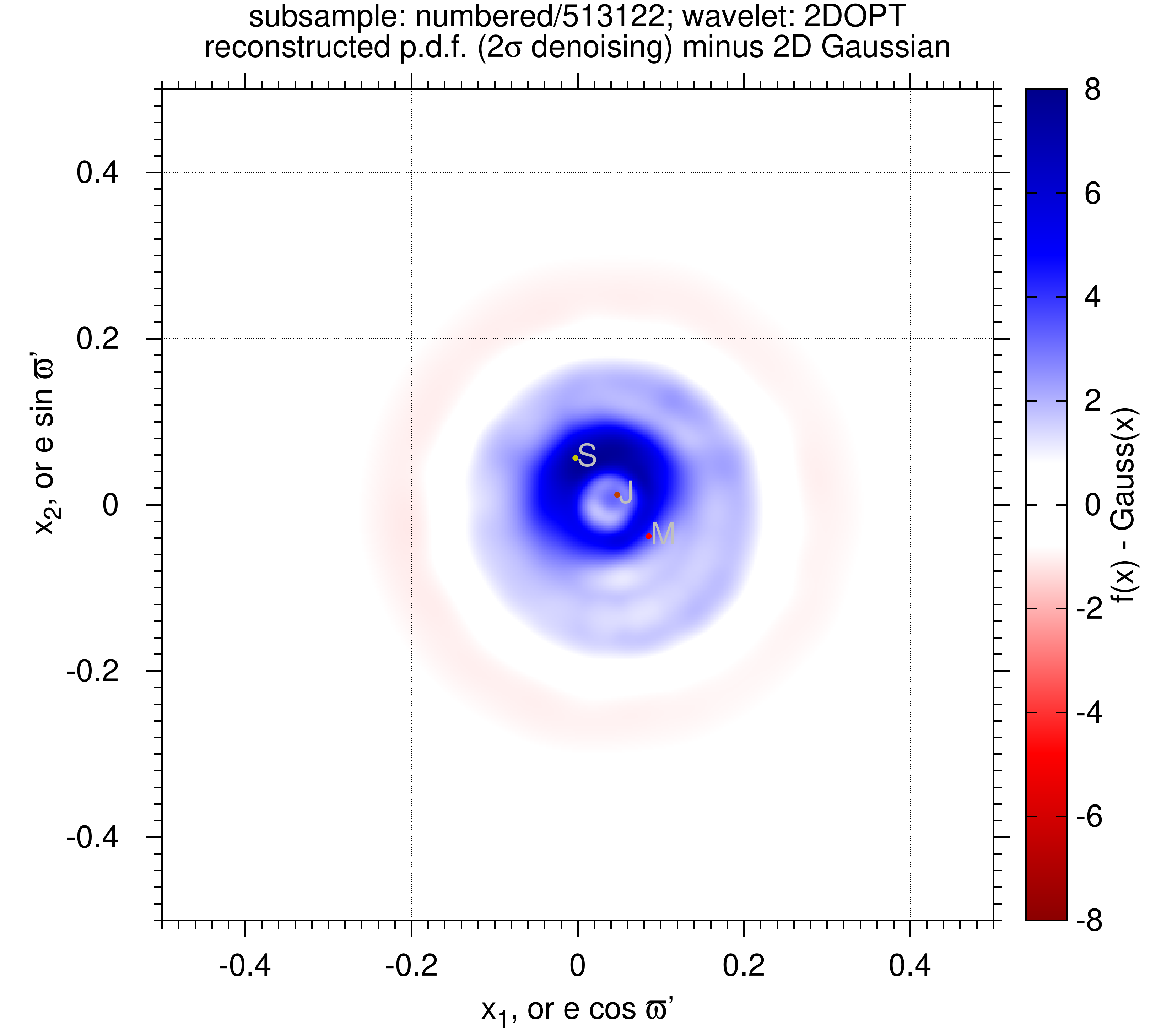}
\end{tabular}
\caption{Reconstructed 2D p.d.f. model $\tilde f(\bmath x)$, and its deviation from
the best fitting radially symmetric Gaussian, based on the CWT analysis from
Fig.~\ref{fig_astecc}.}
\label{fig_astecc_rec}
\end{figure*}

For these parameters we runned our software tool twice. In the first run, we obtained the
CWT at a relatively high resolution $h=0.05$ and with a small minimum scale
$a_{\min}=0.005$. We did not perform the p.d.f. reconstruction in this run, since it
appeared not feasible to complete even a single iteration. In the second run, we increased
the grid step to $h=0.1$ and the minimum scale to $a_{\min}=0.01$. With these options we
were capable to complete about $20$ iterations of the p.d.f. reconstruction phase.

The results are plotted in Fig.~\ref{fig_astecc}. We represent the three-dimensional CWT as
a series of 2D slices taken at different scales $a$. We plot a few of such slices in the
figure, and a video movie showing our results in full is available in the electronic
version of the paper following this figure. Also, the reconstructed p.d.f. model is shown
in Fig.~\ref{fig_astecc_rec}.

The color in the CWT plots encodes the normal quantile corresponding to the value of $z$ in
the given point $(a,\bmath b)$, more explanation given by \citet{Baluev18b}. Notice that
the sign corresponds to the sign of $\Delta f$ (after smoothing), so negative values mean a
convex density function (e.g. near local maximum), while positive values mean concave
p.d.f. (e.g. near local minimum). The hashed domain is where the normality test was failed.

We can see a fascinating evolution of arcing and ringing structures of different scale,
position, and shape. Significant structures remain up to the smallest scales below $a\sim
0.01$, where we can see a small and accurate ring somewhat shifted from the center. The
sign of $z$ indicates a paucity of asteroids along this ring, so it might reflect e.g. some
separatrix in the dynamical phase space (they usually represent sources of chaotic motion
and thus are avoided). We unfortunately have to leave these results without any physical
interpretation, since there is not enough space here. We only notice that wavelet analysis
allows to reveal quite intriguing structures in the distribution, and these results needs
further investigation and assessment in subsequent works.

\section{Conclusions}
Our new version of the algorithm is suitable to analyse 2D distributions, thus it may be
deemed as an improved analogue of the method presented by \citep{Skuljan99}. Let us now
highlight several main improvements of our wavelet analysis method in comparison with
others:
\begin{enumerate}
\item We use a probabilistic detection criterion based on an asymptotic estimation of the
p-value significance (false alarm probability), obtained with an adequate treatment of the
`domain penalty' effect. This approximation is practically accurate and entirely analytic,
thus removing the need of Monte Carlo simulations.
\item We constructed an objective criterion to determine the suitable working domain in the
shift-scale space, based on the Gaussianity requirement.
\item We derived optimal wavelets and optimal reconstruction kernels that allowed to
improve the S/N ratio and to reduce the non-Gaussian deviations, thus expanding the working
domain.
\end{enumerate}

Among the limitations of our algorithm we may note that it is computationally demanding and
hence rather slow. This cannot be avoided to a certain extent, but nevertherless, ways are
possible to further increase the computing speed, which we consider as a work for future.

Another limitation is that our algorithm is entirely isotropic, i.e. it applies only
radially symmetric wavelets with a radially symmetric (i.e. diagonal) scale matrix. This
means that it does not suit well to analyse physically heterogeneous pairs of random
quantities (because they may infer considerably different scales). The extension of the
algorithm to non-isotropic cases is possible, however this results in an additional jump of
dimensionality from $3$ to $5$. This would obviously increase the computational complexity
further, possibly rendering the algorithm impractical.

Further ways to generalize this method may possibly involve a neat reduction of various
statistical distortions, e.g. selection biases and more. This includes, in particular, the
so-called unknown inclination effect in the mass distribution of exoplanets (those detected
by radial velocity technique), and in the stellar rotation velocities. In the both cases,
we actually measure a distorted quantity, $m' = m \sin i$ or $v' = v\sin i$ in place of $m$
or $v$, where $i$ is some inclination angle. The latter inclination is unknown in
individual cases, but its effect on the observed distributions $f(m)$ or $f(v)$ can be
reduced in a statistical sense. It appears that if the inclination $i$ is isotropically
distributed then the p.d.f. for $m$ and for $m'$ become tied by an integral Abel equation.
Solving this integral equation is an ill-posed task, but we believe that our wavelet
analysis technique might be useful to invert that integral in a regularized manner.

Also, our method may be improved to take into account measurement uncertainties in the
input sample (which are currently assumed mathematically accurate).

\section*{Acknowledgements}
This work was supported by the Russian Foundation for Basic Research grant 17-02-00542~A
and by the Presidium of Russian Academy of Sciences programme P-28, subprogramme ``The
space: investigating fundamental processes and their interrelations''.

\appendix

\section{Deriving the general isotropic CWT inversion formula}
\label{sec_ICWT}
Let us seek the inverse wavelet transform for~(\ref{ycwti}) in the form:
\begin{equation}
g(\bmath x) = \int\limits_0^{+\infty} p(\kappa) d\kappa \int\limits_{\mathbb R^n} \Upsilon(\kappa,\bmath c) \gamma\left(\kappa \bmath x + \bmath c\right) d\bmath c,
\label{icwt}
\end{equation}
where $p(\kappa)$ and $\gamma$ are unspecified functions. This formula by definition
assumes radially symmetric wavelets $\psi$. The reconstruction kernel $\gamma$ is also
radially symmetric.

Appling the Fourier transform to~(\ref{ycwti}) and~(\ref{icwt}), we obtain
\begin{equation}
\hat \Upsilon(\kappa,\bmath\omega) = \hat f^*(\kappa \bmath\omega) \hat\psi(\omega), \quad
\hat g(\bmath\omega) = \hat f(\bmath\omega) \hat h(\bmath\omega),
\end{equation}
where
\begin{equation}
\hat h(\omega) = \int\limits_0^{+\infty} \mu\left(\frac{\omega}{\kappa}\right) \frac{p(\kappa) d\kappa}{\kappa^n}, \;
\mu(s) = \hat\gamma(s)\hat\psi^*(s).
\label{hath}
\end{equation}
Our goal is to achieve $\hat h(\omega) \equiv \const$, then we would obtain $\hat g \equiv
C \hat f$.

Remarkably, $\hat h$ appeared radially symmetric as well. Now, perform a replacement
$\kappa \mapsto \omega\kappa$:
\begin{equation}
\hat h(\omega) = \int\limits_0^{+\infty} \mu\left(\kappa^{-1}\right) \frac{p(\omega\kappa)}{\omega^{n-1}\kappa^n} d\kappa.
\label{hath2}
\end{equation}
We may notice that
\begin{equation}
\text{if}\, p(s \kappa)\equiv s^{n-1} p(\kappa) \implies \hat h(\omega) = \const,
\end{equation}
as we aimed to achieve. Therefore, $p(\kappa)$ must be homogeneous of degree $n-1$. Since
it is a function of a single-dimensional argument, it must be simply a power function
$p(\kappa) = \kappa^{n-1}$. This implies $\hat h(\omega)\equiv C_{\psi\gamma}$. Now we only
need to return back to the function originals to write down~(\ref{icwtiso}).

Notice that $\omega=0$ is a peculiar point. From~(\ref{hath}), we have $\hat h(0)=0$ due to
the required mutual admissibility property of $\gamma$ and $\psi$. But all other values of
$\hat h$ are equal to $C_{\psi\gamma}$, by the construction of $h$. This appears because
the Fubini theorem does not actually work when applying the Fourier transform
to~(\ref{icwt}), so we cannot interchange the integrals legally, and~(\ref{hath}) becomes
invalid. The reason is because of the integrand behaviour near $\kappa=0$:
\begin{equation}
p(\kappa) \Upsilon(\kappa,\bmath c) \gamma(\kappa \bmath x + \bmath c) \stackrel{k\to 0}{\sim} \kappa^{n-1} \psi(\bmath c) \gamma(\kappa\bmath x + \bmath c).
\end{equation}
By integrating its absolute value by $\bmath x$, then by $\bmath c$, we obtain
\begin{equation}
\kappa^{-1} \int |\psi(\bmath t)| d\bmath t \int |\gamma(\bmath t)| d\bmath t,
\end{equation}
which cannot be integrated further with respect to $\kappa$, because $\int d\kappa/\kappa$
does not converge near $\kappa=0$. Hence, the absolute convergence is broken and the Fubini
theorem cannot be applied directly, while all the derivation above appears merely symbolic.

To make this derivation more mathematically rigorous, we need to replace the low limit in
the outer integral of~(\ref{icwt}) by some $\kappa_{\min}>0$ and then consider the limit
for $\kappa_{\min}\to 0$. In this case~(\ref{hath}) becomes legal, with the lower limit
replaced by $\kappa_{\min}$, and in~(\ref{hath2}) the lower limit should be
$\kappa_{\min}/\omega$. Finally,
\begin{equation}
\hat h(\omega) = \int\limits_0^{a_{\max}\omega} \mu(s) \frac{ds}{s},
\end{equation}
where $a_{\max}=1/\kappa_{\min}$. As we can see, this $\hat h(\omega)$ is no longer
strictly constant, although for large $a_{\max}$ it is almost constant for $\omega \gg
1/a_{\max}$. As $a_{\max}$ tends to infinity, the limit of $\hat h(\omega)$ is
$C_{\psi\gamma}$ for all $\omega \neq 0$, but the point $\omega = 0$ always remains
special.

Whether or not this pecularity is important, depends on the properties of $f(\bmath x)$.
The inverse wavelet transform may be unable to correctly reconstruct some `infinite-scale'
structure of $f$: constant offset, trends, etc. This is not very surprising, because e.g.
if $\int \psi =0$ than any constant offset, added to $f(\bmath x)$, is basically `killed'
by the CWT, so it cannot be reconstructed by the inversion formula. In fact, any harmonic
function $f_0$, i.e. any solution of the Laplace equation $\Delta f_0(\bmath x) = 0$, would
be `killed' by the CWT, if we choose $\psi = \Delta\varphi$. In this case, the
formula~(\ref{icwt}) may produce some $g(\bmath x) = f(\bmath x) - f_0(\bmath x)$ with an
unknown harmonic $f_0$. What this $g$ could be? By the Liouville theorem, a harmonic
function $f_0$ cannot be bounded (except if it is a constant), but $g$ must be bounded
thanks to the wavelet localization property. So in general the inversion
formula~(\ref{icwt}) should result in a bounded non-harmonic remnant of $f(\bmath x)$. But
if $f(\bmath x)$ is a distribution density then it is necessarily bounded and
well-localized. Then the formula~(\ref{icwt}) would reconstruct it literally, without any
distortions, because such a function cannot contain any `infinite-scale' structure
mentioned above.

We do not give a formal mathematical proof for this conclusion, since mathematical theory
is beyond the scope of this paper, but this subtle matter needs a detailed discussion from
the practical point of view.

For example, if $\hat h(0)=0$ then for any finite $a_{\max}$ the reconstruction
$f_{a_{\max}}(x)$ always integrates to zero. Hence, if the formula~(\ref{icwtiso}) is
understood as a limit for $a_{\max}\to\infty$, would not this limit integrate to zero too?
But $f(x)$ is a probability density, so it must integrate to unit. So what does this
apparent contradiction mean in practice?

In practical tests it appeared that the reconstruction $f_{a_{\max}}(x)$ has very long
(about $a_{\max}$) but small negative tails, while closely following $f(x)$ in the central
part. Whenever $a_{\max}\to\infty$, the negative tails of the reconstruction become longer,
simultaneously approaching zero. In the end, we obtain just $f(x)$ itself. Therefore,
although all our partial reconstructions integrate to zero, their limit for
$a_{\max}\to\infty$ equals to $f(x)$ and integrates to what $f(x)$ integrates, i.e. to
unity. Mathematically, this apparent paradox just means that such convergence is not
uniform, so we formally cannot interchange the integration with taking the limit:
\begin{equation}
\lim_{a_{\max}\to +\infty} \int\limits_{\mathbb R^n} f_{a_{\max}}(\bmath x) d\bmath x \neq \int\limits_{\mathbb R^n} \lim_{a_{\max}\to +\infty} f_{a_{\max}}(\bmath x) d\bmath x.
\end{equation}
So, whenever $f(x)$ is bounded and localized well enough (hence, does not have an
`infinite-scale' structure like non-zero constant level or a trend or other harmonic
remnant), the inversion formula~(\ref{icwtiso}) should work literally, i.e. it should
reconstruct $f$ without any corruptions.

In practice, however, we cannot compute the limit, and always deal with some finite
$a_{\max}$ following from the discretization step of the $\kappa$ variable.\footnote{Our
code implies $a_{\max} \sim 3\sigma/h$.} Therefore, all p.d.f. reconstructions that we are
able to compute from~(\ref{icwtiso}) numerically, would integrate to zero and would then
have small but long negative tails. Formally, such a behaviour is non-physical, because the
probability cannot be negative. However, in practice these negative tails remain small in
absolute magnitude, so they just can be neglected. In fact, this is done automatically if
$f(x)$ is computed in the range much smaller than $a_{\max}$. Moreover, the remaining
`kernel' part of the reconstruction automatically appears to integrate close to unit.

\section{2D wavelet transform of radially symmetric Gaussian and flat distributions}
\label{sec_gf}
Using the definition~(\ref{ycwti}), let us substitute a radially symmetric p.d.f.:
\begin{equation}
Y(a, \bmath b) = \int\limits_{\mathbb R^n} f(x) \psi\left(\frac{|\bmath x-\bmath b|}{a}\right) d\bmath x.
\end{equation}

Applying a replacement $\bmath t = (\bmath x - \bmath b)/a$ and using spherical
parametrization, this can be rewritten in two equivalent forms,
\begin{equation}
Y(a, b) = a^n \int\limits_0^{+\infty} \psi(t) t^{n-1} dt \oint f\left(\sqrt{a^2 t^2 + b^2 + 2 a b t \cos\theta}\right) d\Omega,
\end{equation}
where $\theta$ is the angle between $\bmath t$ and $\bmath b$, while $\Omega$ is the solid
angle in $\mathbb R^n$, which is integrated over the entire unit sphere $\mathcal S_n$.
Notice that $Y$ now depends only on the scalar arguments $a$ and $b$, meaning it became
radially symmetric with respect to $\bmath b$.

If $f(x)$ is a radially-symmetric flat distribution then we obtain
\begin{align}
f(x) &= \left\{ \frac{1}{S_n}, \, x\leq 1, \atop 0, \, x>1, \right. \implies \nonumber\\
Y(a, b) &= a^n \int\limits_0^{+\infty} \psi(t) t^{n-1} dt \int\limits_{a^2 t^2 + b^2 + 2 a b t \cos\theta \leq 1} \frac{d\Omega}{S_n},
\end{align}
The condition in the inner integral is equivalent to a condition on the angle $\theta$ in
the form
\begin{equation}
\cos\theta \leq \frac{1-b^2-a^2 t^2}{2abt},
\label{cost}
\end{equation}
which restrict some general solid angle, or a domain on the unit sphere. This domain
depends on $a$, $b$, and $t$, and for some their values this domain may turn either empty
or cover the complete sphere, because the right hand side of~(\ref{cost}) may become
smaller than $-1$ or above $+1$, respectively. So implicitly this inequality may also
restrict the integration domain for $t$.

Notice that all of $a$, $b$, and $t$ are non-negative numbers here. Using this property,
after some elemetary manipulations we can set the following constraint on $\theta$
explicitly:
\begin{align}
\theta &\in [\pi-\theta_0, \pi+\theta_0], &t \in\left[\frac{|b-1|}{a}, \frac{b+1}{a} \right],\nonumber\\
& & \theta_0 = \arccos \frac{b^2+a^2 t^2-1}{2abt}, \nonumber\\
\theta &\in [0, 2\pi], &t \in \left[0, \frac{1-b}{a}\right], \, b<1,
\end{align}
and empty set otherwise.

Therefore,
\begin{align}
a^{-n} Y(a, b) &= \int\limits_0^{\frac{1-b}{a}} \psi(t) t^{n-1} dt + \int\limits_{\frac{|b-1|}{a}}^{\frac{b+1}{a}} \psi(t) t^{n-1} dt \int\limits_{|\pi-\theta|\leq \theta_0} \frac{d\Omega}{S_n},
\end{align}
where the first integral should be removed, if $b>1$. In the 2D case, the integration with
respect to $d\Omega$ is equivalent to integration over just $\theta$, so the formula becomes
\begin{equation}
Y(a, b) = a^2 \int\limits_0^{\frac{1-b}{a}} \psi(t) t dt + \frac{a^2}{\pi} \int\limits_{\frac{|b-1|}{a}}^{\frac{b+1}{a}} \theta_0(t) \psi(t) t dt,
\end{equation}
Now, since $\psi(\bmath t) = \Delta\varphi(\bmath t)$, by the radial symmetry we have
$\psi(t) = t^{-n} (t^n\varphi'(t))'$, and hence $\psi(t) t^{n-1} dt = d(t^n\varphi'(t))/t$.
This allows us to derive the following identity:
\begin{align}
Q_n(t) &:= \int q(t) \psi(t) t^{n-1} dt = \int \frac{q(t)}{t} d(t^n\varphi'(t)) \nonumber\\
&= q \varphi' t^{n-1} - \int \left(\frac{q(t)}{t}\right)' t^n d\varphi(t) \nonumber\\
&= q \varphi' t^{n-1} - \int \left(q' t^{n-1} - q t^{n-2}\right) d\varphi \nonumber\\
&= q \varphi' t^{n-1} - \int q'(t) \varphi' t^{n-1} dt + \int q t^{n-2} d\varphi \nonumber\\
&= (t\varphi' + \varphi) q t^{n-2} - \int q' \varphi' t^{n-1} dt - \int \left(q t^{n-2}\right)' \varphi(t) dt \nonumber\\
&= (t\varphi' + \varphi) q t^{n-2} - \int \left[(t \varphi' +\varphi) q' + \frac{n-2}{t} q \varphi \right] t^{n-2} dt.
\end{align}
For $n=2$ this remarkably simplifies to just
\begin{equation}
Q_2(t) = (t\varphi' + \varphi) q  - \int (t\varphi' + \varphi) q' dt,
\end{equation}
so the integration can be also understood with respect to $q$ as independent variable
(expressing $t$ via $q$).

By substituting $q=1$ and $q=\theta_0(t)$, we obtain:
\begin{align}
a^{-2} Y(a, b) = \left.\left(t\varphi'+\varphi\right)\right|_{t=\max(\frac{1-b}{a},0)} - \varphi(0) \nonumber\\
+ \frac{1}{\pi} \left. \left[ (t\varphi'+\varphi) \theta_0 \right] \right|_{\frac{|b-1|}{a}}^{\frac{b+1}{a}}
- \frac{1}{\pi}\int\limits_{\frac{|b-1|}{a}}^{\frac{b+1}{a}} (t\varphi'+\varphi) \theta_0' dt.
\end{align}
Now we may use the property that $\theta_0=0$ at the right endpoint, and $\theta_0$ at the
left endpoint is either $0$, if $b>1$, or $\pi$, if $b<1$. Moreover, we can now apply a
replacement of the integration variable $t \mapsto \theta_0$, so we have:
\begin{align}
Y(a, b) &= \left\{ -a^2 \varphi(0) + \frac{a^2}{\pi}\int\limits_0^{\pi} \left. (t\varphi'+\varphi)\right|_{t=t_+(\theta)} d\theta, \quad b<1,
 \atop \frac{a^2}{\pi} \int\limits_0^{\arcsin\frac{1}{b}} \left.(t\varphi'+\varphi)\right|_{t_-(\theta)}^{t_+(\theta)} d\theta, \quad b>1, \right. \nonumber\\
t_\pm(\theta) &= \frac{1}{a}\left( b\cos\theta \pm \sqrt{1-b^2\sin^2\theta} \right).
\end{align}
These integrals can be computed numerically (we use the GNU scientific library for that).

Notice that if $b>1$, one may mistakenly assume that in terms of $\theta$ the length of the
integration segment becomes zero and hence the integral vanishes. This is wrong, because
$\theta_0'$ changes sign at $t_*=\sqrt{b^2-1}/a$, if $b>1$. We should split the integration
segment into two portions, separated by this change point, and compute these integrals
independently, assuming different solution branches for $t$, either $t_1(\theta)$ or
$t_2(\theta)$.

Now let us consider the case of standard Gaussian density $f(x)$:
\begin{align}
f(x) &= \frac{1}{(2\pi)^{\frac{n}{2}}} e^{-\frac{x^2}{2}} \implies \nonumber\\
Y(a, b) &= a^n \int\limits_0^{+\infty} \psi(t) t^{n-1} dt \oint e^{-\frac{a^2 t^2+b^2}{2} - a b t \cos\theta} \frac{d\Omega}{(2\pi)^{\frac{n}{2}}},
\end{align}
which for $n=2$ turns into
\begin{equation}
Y(a, b) = a^2 e^{-\frac{b^2}{2}} \int\limits_0^{+\infty} e^{-\frac{a^2 t^2}{2}} \psi(t) t dt \int\limits_0^{2\pi} e^{- a b t \cos\theta} \frac{d\theta}{2\pi}.
\end{equation}
The innermost integral can be expressed via the modified Bessel function $I_0$, so we have
\begin{align}
Y(a, b) &= a^2 e^{-\frac{b^2}{2}} \int\limits_0^{+\infty} e^{-\frac{a^2 t^2}{2}} I_0(a b t) \psi(t) t dt \nonumber\\
 &= a^2 \int\limits_0^{+\infty} e^{-\frac{(a t-b)^2}{2}} \left[e^{-abt} I_0(a b t)\right] \psi(t) t dt
\label{Ygauss}
\end{align}
The last formula was necessary to replace $I_0$ in the integrand by a normalized form
$e^{-z} I_0(z)$, which is a bounded well-behaved function decreasing as $\sim 1/\sqrt z$ in
the tail. This integral can be computed numerically.

In the both cases, we could not find a fully analytic representation of the wavelet
transform, except for the 2DMHAT wavelet, for which the integral~(\ref{Ygauss}) was
computed by means of computer algebra:
\begin{equation}
Y(a,b) = \left(\frac{a^2}{a^2+1}\right)^2 \left( 2 - \frac{b^2}{a^2+1} \right) e^{-\frac{1}{2}\frac{b^2}{a^2+1}}.
\end{equation}

\bibliographystyle{model2-names}
\bibliography{wavelets}







\end{document}